\documentstyle[preprint,tighten,floats,prd,aps]{revtex}
\def \pom {{\hspace{ -0.1em}I\hspace{-0.2em}P}}
\input epsf.tex
\def\DESepsf(#1 width #2){\epsfxsize=#2 \epsfbox{#1}}
\begin{document}
\draft
%
\preprint{\vbox{
\hbox{PSU/TH/163}
\hbox{OITS 581}
\hbox{hep-ph/9509239}}  }
\title{
Behavior of Diffractive Parton Distribution Functions}
\author{ Arjun Berera}
\address{
Department of Physics\\
Pennsylvania State University,
University Park, PA 16802
}
\author{ Davison E.\ Soper}
\address{
Institute of Theoretical Science\\
University of Oregon,
Eugene, OR  97403
}
\date{5 September 1995}
\maketitle
\begin{abstract}
Diffractive parton distribution functions give the probability to
find a parton in a hadron if the hadron is diffractively scattered.
We provide an operator definition of these functions and discuss their
relation to diffractive deeply inelastic scattering and to
photoproduction of jets at HERA. We perform a calculation in the
style of ``constituent counting rules'' for the behavior of these
functions when the detected parton carries almost all of the
longitudinal momentum transferred from the scattered hadron.
\end{abstract}
\narrowtext
%
%

\section{Introduction}
\label{Intro}

Recently, the Zeus and H1 experiments at HERA have reported the first
evidence for diffractive deeply inelastic electron scattering
\cite{ZeusH1},
\begin{equation}
e + A \to e + A^\prime + X.
\label{ddis}
\end{equation}
This is an example of a more general phenomenon, diffractive hard
scattering, in which a high energy incident hadron participates in a
hard interaction, involving very large momentum transfers, but
nevertheless the hadron itself is diffractively scattered, emerging
with a small transverse momentum and the loss of a rather
small fraction of its longitudinal momentum. One may say that the
hadron has exchanged a pomeron with the rest of the particles
involved and that the pomeron has participated in the hard
interaction. The possibility of such interactions was proposed by
Ingelman and Schlein \cite{IS} on the grounds that the entity
exchanged in elastic scattering, called the pomeron, must be made of
quarks and gluons, which, being pointlike, can participate in hard
interactions. The theoretical ideas and formulas involved are
elaborated in some detail in Ref.\ \cite{BCSS}. The predicted
phenomenon was seen in jet production in hadron collisions by the UA8
Collaboration \cite{UA8}.

As discussed in our previous work \cite{BS}, the Ingelman-Schlein
model can be thought of as involving a ``diffractive parton
distribution function,'' which is the subject of this paper.
The idea is that this function,
\begin{equation}
{d\, f^{\rm diff}_{a/A}(\xi,\mu;x_\pom,t)\over dx_\pom\,dt}
\label{fdiff}
\end{equation}
represents, in a hadron of type $A$, the probability per unit $d\xi$
to find a parton of type $a$ carrying momentum fraction $\xi$, while
leaving hadron $A$ intact except for the momentum transfer
characterized by parameters $(x_\pom,t)$. Here $t$ is the invariant
momentum transfer $t = (P_{\!A} - P_{\!A^\prime})^2$ while
$x_\pom$ is the fraction of its original longitudinal momentum lost
by the hadron. The parameter $\mu$ is the factorization scale,
roughly, the resolution of the parton probe. A function expressing
the same physics as the diffractive parton distribution (\ref{fdiff})
has been proposed by Veneziano and Trentadue \cite{VT} under the name
of ``fracture function.'' The details are a little different, as we
will explain in Sec.~\ref{DiffDIS}. The original paper of Ingelman and
Schlein did not mention the function (\ref{fdiff}) but instead
introduced a related function, the ``distribution of partons in the
pomeron.''

Our purpose in this paper is, first of all, to relate these various
functions and the ideas behind them to one another and to comment on
the likely validity of the formulas that express cross sections in
terms of these functions. We give operator definitions for the
diffractive parton distribution functions and discuss the evolution
equation that they obey. We briefly review the expected behavior
of ${d\, f^{\rm diff}_{a/A}(\xi,\mu;x_\pom,t)/ dx_\pom\,dt}$ for small
$\beta$, where $\beta = \xi/x_\pom$. Then we use a perturbative
calculation to explore how these functions behave for small values of
$1-\beta$. In the Ingelman-Schlein language, our results favor a
rather ``hard'' distribution of partons in the pomeron. We also
present, in an appendix, a calculation of ${d\, f^{\rm
diff}_{a/A}(\xi,\mu;x_\pom,t)/ dx_\pom\,dt}$ in a simple model. We
conclude with some observations on the experimental consequences of
the theory.

\section{Diffractive deeply inelastic scattering}
\label{DiffDIS}

\begin{figure}[htb]
\centerline{ \DESepsf(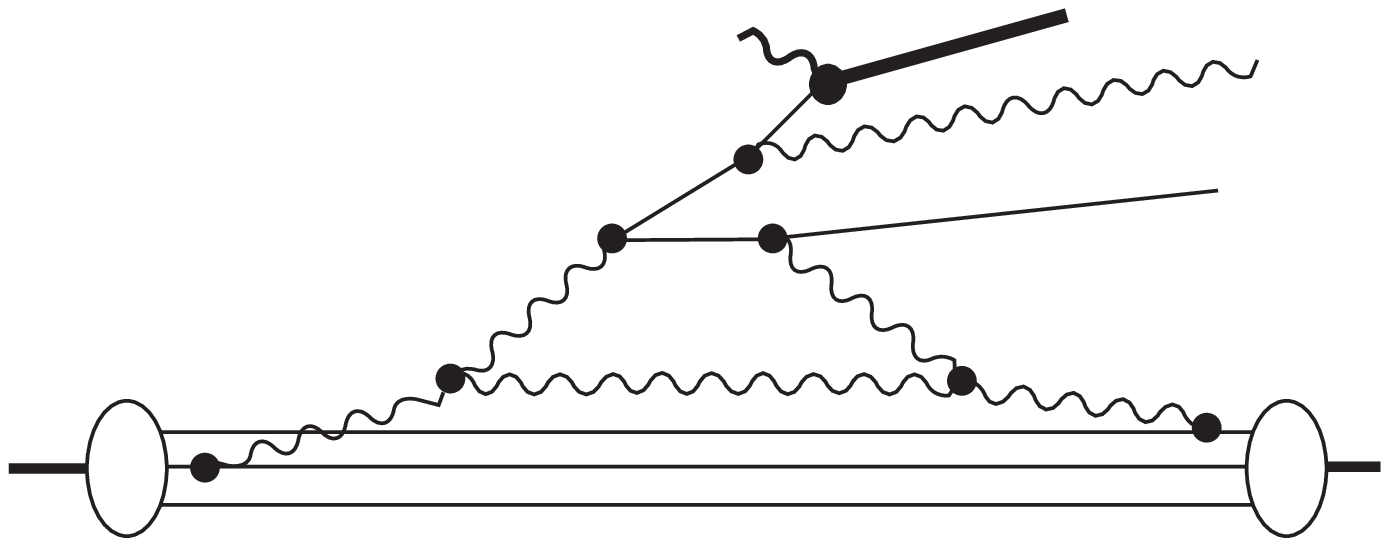 width 10 cm) }
\vskip 0.5 cm
\caption{
A graph for $e + p \to p + X$.}
\label{diffDIS}
\end{figure}

In a deeply inelastic scattering reaction, a hadron $A$ with momentum
$P_{\!A}^\mu$ is struck by a far off shell photon with momentum
$q^\mu$. It is convenient to use momentum components $k^\mu =
(k^+,k^-,{\bf k})$, where $k^\pm = 2^{-1/2}(k^0 \pm k^3)$, and where
we denote transverse components of vectors by boldface. We work in
the brick wall frame, in which $P_{\!A}^\mu = (P_{\!A}^+,
M_{\!A}^2/[2P_{\!A}^+], {\bf 0})$ and $q^\mu = 2^{-1/2}(-Q,Q,{\bf
0})$.  One measures the standard hard scattering variables $Q^2 =-
q\cdot q$  and $x = Q^2/[2 P_{\!A}\cdot q$]. In some deeply inelastic
scattering events there will be in the final state a diffractively
scattered hadron $A^\prime$ with momentum
\begin{equation}
P_{\!A^\prime} = \left(
[1-x_\pom]P_{\!A}^+,{{\bf P}_{\!A^\prime}^{2} + M_{\!A}^2\over
2[1-x_\pom]P_{\!A}^+}, {\bf P}_{\!A^\prime}\right)
\label{PAprime}
\end{equation}
as in  Fig.~(\ref{diffDIS}). The hadron has lost a fraction $x_\pom$
of its plus momentum and has gained transverse momentum ${\bf
P}_{\!A^\prime}$. The invariant momentum transfer from the proton, $t
=  (P_{\!A} - P_{\!A^\prime})^2$, is
\begin{equation}
t = - {{\bf P}_{\!A^\prime}^{2}+ x_\pom^2 M_{\!A}^2 \over 1-x_\pom }.
\label{t}
\end{equation}
The events in which we are interested have small $t$. One expects
$|t| \lesssim 1\ {\rm GeV}^2$ to be typical. We also suppose that
$x_\pom$ is rather small. One expects pomeron physics to be dominant
for $x_\pom<0.1$. Having found such events, one can construct the
contribution to $F_2$ from final states containing a diffractively
scattered hadron with variables $t$ and $x_\pom$:  $dF_2^{\rm
diff}(x,Q^2;x_\pom,t)/ dx_\pom\,dt$.

The model of Ingelman and Schlein \cite{IS} as applied to deeply
inelastic scattering is simple to state. We begin with the usual
factorization theorem for the structure function $F_2$:
\begin{equation}
F_2(x,Q^2) = \sum_a \int_0^1 d\xi\ f_{a/A}(\xi,\mu)\
\hat F_{2,a}(x/\xi,Q^2;\mu).
\end{equation}
Here $f_{a/A}(\xi,\mu)$ is the distribution of partons of kind $a$
in hadron $A$ as a function of momentum fraction $\xi$, as measured at
a factorization scale $\mu$, while $\hat F_{2,a}$ is the structure
function for deeply inelastic scattering on parton $a$. If, for
simplicity, we ignore $Z$ exchange, then $\hat F_2$ is
\begin{equation}
\hat F_{2,a}(x/\xi,Q^2;\mu)
= e_a^2\ \delta(1-x/\xi) + {\cal O}(\alpha_s).
\end{equation}
Thus $F_2$ is rather trivially related to the parton distribution
functions at the Born level; nevertheless, conceptually the
distinction between $F_2(x,Q^2)$ and $f_{a/A}(\xi,\mu)$ is quite
important. As in our previous paper \cite{BS}, we break the analysis
into two stages.  In the first stage, we hypothesize that the
diffractive structure function $F_2^{\rm diff}$ can be written in
terms of a diffractive parton distribution, Eq.~(\ref{fdiff}):
\begin{equation}
{d F_2^{\rm diff}(x,Q^2;x_\pom,t)\over dx_\pom\,dt}
= \sum_a \int_0^{x_\pom} d\xi\
{d\, f^{\rm diff}_{a/A}(\xi,\mu;x_\pom,t)\over dx_\pom\,dt}\
\hat F_{2,a}(x/\xi,Q^2;\mu).
\label{factor}
\end{equation}
In the second stage, we hypothesize that ${d\, f^{\rm
diff}_{a/A}(x_a,\mu)/ dx_\pom\,dt}$ has a particular form:
\begin{equation}
{d\,f^{\rm diff}_{a/A}(\xi,\mu;x_\pom,t)\over dx_\pom\,dt}
={1 \over 8\pi^2}\, |\beta_{A}(t)|^2 x_\pom^{-2\alpha(t)}\,
f_{a/\pom}(\xi/x_\pom, t,\mu)\,.
\label{pomdist}
\end{equation}
Here $\beta_A(t)$ is the pomeron coupling to hadron A and $\alpha(t)$
is the pomeron trajectory. We distinguish the ``Regge factorization''
of Eq.~(\ref{pomdist}) from the ``diffractive factorization'' of
Eq.~(\ref{factor}).

In Eq.~(\ref{pomdist}) we adopt standard conventions such
that the proton-proton elastic scattering amplitude is
\begin{equation}
{\cal M} = - \beta_p(t)^2 s^{\alpha(t)}\,.
\label {ppelas}
\end{equation}
Then the elastic scattering cross section is
\begin{equation}
{d\sigma \over dt}
= {1 \over 16 \pi} |\beta_p(t)|^4 s^{2(\alpha(t) - 1)}\,.
\end{equation}
while the total proton-proton cross section is
\begin{equation}
\sigma_{\rm tot}(pp)
= \Re [ \beta_p(0)^2]\, s^{\alpha(0)-1}\,.
\end{equation}
The normalization factor $1/(8\pi^2)$ in Eq.~(\ref{pomdist}) is quite
arbitrary. Here, we have adopted the convention of Donnachie and
Landshoff \cite{DonnachieLandshoff}.

The function $f_{a/\pom}(\beta, t,\mu)$ thus defined is
the ``distribution of partons in the pomeron.'' In writing
Eq.~(\ref{pomdist}), one thinks of the pomeron as a continuation in
the angular momentum plane of a set of hadron states.  Since hadrons
contain partons, the pomeron should also. Thus one has in
Eq.~(\ref{pomdist}) the standard factors describing the coupling of
the pomeron to hadron A, together with a distribution of partons in
the pomeron \cite{IS,BCSS}. Inserting Eq.~(\ref{pomdist}) into
(\ref{factor}), one obtains the model of Ingelman and Schlein,
applied to the case of deeply inelastic scattering:
\begin{equation}
{d F_2^{\rm diff}(x,Q^2;x_\pom,t)\over dx_\pom\,dt}
=
{|\beta_{A}(t)|^2 \over 8\pi^2}\,  x_\pom^{1-2\alpha(t)}\,
 \sum_a \int_0^1 d\tilde\beta\,
f_{a/\pom}(\tilde\beta, t,\mu)\,
\hat F_{2,a}(\beta/\tilde\beta,Q^2;\mu),
\label{IS}
\end{equation}
where $\beta = x/x_\pom$. We offer here a word of caution. Both the
structure of Eq.~(\ref{IS}) and the language ``distribution of
partons in the pomeron'' suggest that the hadron emits a pomeron some
long time before the hard interaction and that the pomeron then
splits into partons, one of which participates in the hard
interaction. This interpretation is, however, not required by
Eq.~(\ref{IS}) and is surely quite misleading. In a diagrammatic
interpretation of pomeron exchange \cite{BFKLgraphs}, the exchanged
quanta have small plus and minus components of momentum. Thus the
exchange takes place over a long interval $(\Delta x^+, \Delta x^-)$
in space-time. It begins long before the hard interaction and ends
long afterwards. Our diagrammatic analysis in Secs.~\ref{TauGlue} and
\ref{TauQuark} will provide an illustration of this picture.

We see that  Eq.~(\ref{factor}), can be regarded as a version of the
Ingelman-Schlein model (\ref{IS}) that is more parsimonious in its
assumptions. Eq.~(\ref{factor}) says only that factorization still
applies when hadron $A$ is diffractively scattered. The
Ingelman-Schlein model (\ref{IS}) assumes that Regge phenomenology is
applicable and, with the aid of this assumption, has more predictive
power.

In this paper, we concentrate on the case in which  hadron $A^\prime$
is the same kind of hadron as hadron $A$, so that vacuum quantum
numbers are exchanged, and we consider $x_\pom$ to be small enough so
that pomeron exchange dominates. One should keep in mind, however,
that Eq.~(\ref{factor}) admits generalizations to cases where
$A^\prime \ne A$ and where $x_\pom$ is not at all small. One can also
generalize Eq.~(\ref{IS}) to $A^\prime \ne A$, but then $x_\pom$
should be fairly small in order that just one or two Regge exchanges
dominate.

The diffractive factorization equation~(\ref{factor}), or rather a
very closely related equation, has been introduced by Veneziano and
Trentadue \cite{VT}.  These authors call the analogue of
${d\,f^{\rm diff}_{a/A}(\xi,\mu)/ dx_\pom\,dt}$ a ``fracture
function.'' Stated precisely, a fracture function is
\begin{equation}
{d\,f^{\rm diff}_{a/A}(\xi,\mu;x_\pom)\over dx_\pom}
= \int_0^\infty d|t|\ {d\,f^{\rm diff}_{a/A}(\xi,\mu;x_\pom,t)
\over dx_\pom\,dt}.
\end{equation}
By integrating over $t$, Veneziano and Trentadue eliminate a
variable that is perhaps of secondary importance. However, there is
some advantage to {\it not} integrating over $t$.   We are interested
in the physics of diffraction, which occurs in the small $t$ region.
If we integrate over $t$, then we are forced to consider also the
large $t$ region, in which the hadron $A^\prime$ is to be thought of
not as the original hadron appearing in a scattered form but as a
random hadron in a high $P_T$ jet produced in the hard interaction.
Taking this possibility into account leads to certain complications
in the formulas.

In the following sections, we analyze the diffractive parton
distributions. According to Eq.~(\ref{factor}), the measured
quantity $dF_2^{\rm diff}/dx_\pom\,dt$ is approximately the sum of
diffractive quark distributions weighted by the square of the quark
charges. There are higher order corrections to this relation, some
involving the diffractive gluon distribution. Thus these
distributions are rather directly related to experiment. The reader
may wonder why we concentrate on the theoretical diffractive parton
distributions rather than on the physical quantity
$dF_2^{\rm diff}/dx_\pom\,dt$. The reasons are the same as in
ordinary hard scattering: 1) the diffractive parton distributions
are process independent and 2) the factorization (\ref{factor})
allows one to include perturbative corrections to the hard
scattering. The reader may also wonder why we don't frame the
analysis in terms of the distribution of partons in the pomeron. Our
excuse is ignorance. We don't know how to relate the Regge
factorization in Eq.~(\ref{pomdist}) to quantum field theory.

\section{The diffractive parton distribution}
\label{DiffPDF}

In this section we give an operator definition of the diffractive
parton distribution. We write the ordinary distribution of a quark of
type $j \in \{u,\bar u,d,\bar d,\dots\}$ in a hadron of type $A$
in terms of field operators $\tilde\psi(y^+,y^-,{\bf y})$
evaluated at $y^+ = 0$, ${\bf y} =0$ \cite{CSdistfns,CFP}:
\begin{eqnarray}
f_{j/A}(\xi,\mu)
&\equiv&\!
{1 \over 4\pi} {1 \over 2}\sum_{s_{\!A}}\! \int\! d y^-
e^{-i\xi P_{\!A}^+ y^-} \langle P_{\!A},s_{\!A} |
\tilde{\overline \psi}_j(0,y^-,{\bf 0})\gamma^+
\tilde\psi_j(0)
| P_{\!A},s_{\!A} \rangle.
\label{quarkdist}
\end{eqnarray}
Similarly, the ordinary distribution of a gluon in a proton
is written as
\begin{eqnarray}
f_{g/A}(\xi,\mu)
&\equiv&\!
{1 \over 2\pi \xi P_{\!A}^+} {1 \over 2}\sum_{s_{\!A}}\! \int\! d y^-
e^{-i\xi P_{\!A}^+ y^-} \langle P_{\!A},s_{\!A} |
\tilde F_a^\dagger(0,y^-,{\bf 0})^{+\nu}
\tilde F_a(0)_\nu^{\ +}
|P_{\!A},s_{\!A} \rangle.\ \
\label{gluedist}
\end{eqnarray}
The proton state $| P_{\!A},s_{\!A} \rangle$ has spin $s_{\!A}$ and
momentum $P_{\!A}^\mu =  (P_{\!A}^+, {M_{\!A}^2 / [2P_{\!A}^+]},{\bf
0})$. We average over the spin. Our states are normalized to
\begin{equation}
\langle k |p \rangle = (2\pi)^3\,
2p^0\,\delta^3(\vec p - \vec k) =
 (2\pi)^3\, 2p^+\,\delta(p^+ - k^+)\,\delta^2({\bf p} - {\bf k})
\end{equation}
The field $\tilde \psi_j(0,y^-,{\bf 0})$ is the quark field
operator modified by multiplication by an exponential of a line
integral of the vector potential:
\begin{equation}
\tilde \psi_j(0,y^-,{\bf 0})
=
\left[
{\cal P}
\exp\left(
i g \int_{y^-}^\infty d x^-\, A_c^+(0,x^-,{\bf 0})\, t_c
\right)
\right]
\psi_j(0,y^-,{\bf 0})\,.
\label{tildepsi}
\end{equation}
Likewise $\tilde F_a(0,y^-,{\bf 0})^{+\nu}$ is defined by
\begin{equation}
\tilde F_a(0,y^-,{\bf 0})^{\mu\nu}
=
\left[
{\cal P}
\exp\left(
i g \int_{y^-}^\infty d x^-\, A_c^+(0,x^-,{\bf 0})\, t_c
\right)
\right]_{ab}
F_b(0,y^-,{\bf 0})^{\mu\nu}\,.
\label{tildeF}
\end{equation}
The ${\cal P}$ denotes path ordering of the exponential. The matrices
$t_c$ in Eq.~(\ref{tildepsi}) are the generators of the $\bf 3$
representation of SU(3), while in Eq.~(\ref{tildeF}) they are the
generators of the $\bf 8$ representation. These operator products are
ultraviolet divergent, and are renormalized at scale $\mu$ using the
$\overline{\rm MS}$ prescription, as described in
\cite{CSdistfns}.

The motivation for these definitions is that in QCD canonically
quantized on null surfaces $x^+ = const.$ using $A^+ = 0$ gauge, the
operators measure the probability to find a quark and gluon
respectively carrying plus component of momentum equal to
$\xi P_{\!A}^+$. The line integrals of the color potential restore gauge
invariance. Then $\overline{\rm MS}$ renormalization removes
divergences.

The line integrals of the color potential have a physical
interpretation. Whenever a parton is measured by a short distance
probe, the color carried by that parton has to go somewhere. For
instance, in deeply inelastic scattering, the color is carried away
by the recoiling struck quark. In the definition of the parton
distribution function, the recoil color flow is idealized as an
infinitely narrow jet moving with the speed of light along the path
$x^\mu = (0,x^-,{\bf 0})$ with $y^- < x^- <\infty$. Any gluons from
the color field of the hadron can couple to this idealized color
source.

Consider now the diffractive distribution of a quark in a proton. The
operator is the same as in Eq.~(\ref{quarkdist}), but the proton  is
required to appear in the final state carrying momentum $P_{\!A}^\prime$:
\begin{eqnarray}
\lefteqn{(2 \pi)^3 2 E_{\!A^\prime}\
{d\, f^{\rm diff}_{j/A}(\xi,\mu)\over
d^3 \vec P_{\!A^\prime}}
= G_{j/A}^{\rm diff}(P_{\!A},p_{\!A^\prime},\xi,\mu)}
\nonumber\\
&\equiv&
{1 \over 4\pi}{1 \over 2}\sum_{s_{\!A}}\int d y^- e^{-i\xi P_{\!A}^+ y^-}
\sum_{X,s_{\!A^\prime}}
\langle P_{\!A},s_{\!A} |\tilde {\overline\psi}_j(0,y^-,{\bf 0})
| P_{\!A^\prime},s_{\!A^\prime}; X \rangle
\nonumber\\
&& \hskip 3.5 cm  \times
\gamma^+ \langle P_{\!A^\prime},s_{\!A^\prime}; X|
{\tilde {\psi}}_j(0)| P_{\!A},s_{\!A} \rangle.
\label{fdiff1}
\end{eqnarray}
We sum over the spin $s_{\!A^\prime}$ of the final state proton and over
the states $X$ of any other particles that may accompany it.
Similarly, the diffractive distribution of gluons in a hadron is
\begin{eqnarray}
\lefteqn{(2 \pi)^3 2 E_{\!A^\prime}\
{d\, f^{\rm diff}_{g/A}(\xi,\mu)\over
d^3 \vec P_{\!A^\prime}}
= G_{g/A}^{\rm diff}(P_{\!A},P_{\!A^\prime},\xi,\mu)}
\nonumber\\
&\equiv&
{1 \over 2\pi \xi P_{\!A}^+}{1 \over 2}\sum_{s_{\!A}} \int d y^-
e^{-i\xi P_{\!A}^+ y^-}
\sum_{X,s_{\!A^\prime}}
\langle P_{\!A},s_{\!A} |\tilde F_a(0,y^-,{\bf 0})^{+\nu}
| P_{\!A^\prime},s_{\!A^\prime}; X \rangle
\nonumber\\
&& \hskip 4 cm  \times
\langle P_{\!A^\prime},s_{\!A^\prime}; X|
\tilde F_a(0)_\nu^{\ +}| P_{\!A},s_{\!A} \rangle.
\label{fdiff2}
\end{eqnarray}
The Green function $G_{a/A}^{\rm diff}$ for a parton of type $a$ can,
at least in principle, be computed from Feynman diagrams together with
the Bethe-Salpeter wave functions for the bound states.

We will want to change variables to $x_\pom$ and $t$ as defined in the
previous section. Using Eqs.~(\ref{PAprime}) and (\ref{t}), we
obtain
\begin{equation}
{d^3 {\vec P_{\!A^\prime}} \over 2 E_{\!A^\prime}}
= {d x_\pom \over 2(1-x_\pom)}\  d^2 {\bf P}_{\! A^\prime}
= {1 \over 4}\, dx_\pom\,dt \, d\phi\,.
\end{equation}
Integrating over the azimuthal angle $\phi$, we have
\begin{equation}
{d\, f^{\rm diff}_{a/A}(\xi,\mu)\over
dx_\pom\,dt}
={1 \over 16\pi^2}\ G_{a/A}^{\rm diff}(P_{\!A},P_{\!A^\prime},\xi,\mu)
\label{fdifftoG}
\end{equation}
where $G_{j/A}^{\rm diff}$ for quarks is given in Eq.~(\ref{fdiff1})
and $G_{g/A}^{\rm diff}$ for gluons is given in Eq.~(\ref{fdiff2}).

\section{Evolution equation}
\label{Evolution}

As mentioned in the previous section, the diffractive parton
distributions are ultraviolet divergent and require
renormalization. It is convenient to perform the renormalization
using the $\overline{\rm MS}$ prescription, as
discussed in \cite{CSdistfns,CFP}. This introduces a renormalization
scale $\mu$ into the functions. In applications, one sets $\mu$ to be
the same order of magnitude as the hard scale of the physical process.

\begin{figure}[htb]
\centerline{ \DESepsf(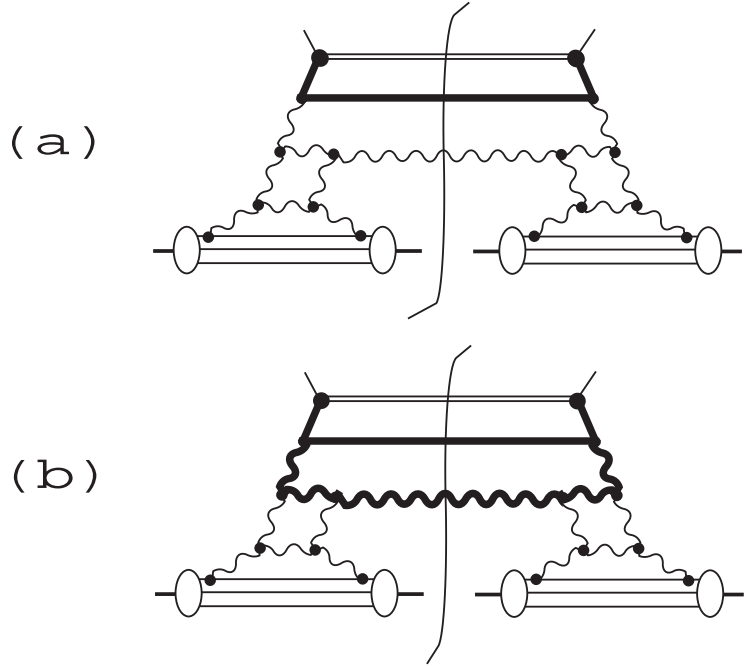 width 10 cm) }
\vskip 0.5 cm
\caption{
Renormalization of the diffractive quark distribution. The subgraph
in (a) denoted by heavy lines is ultraviolet divergent, and thus
contributes to the evolution kernel. The subgraph in (b) is not
ultraviolet divergent (with transverse polarizations for the
incoming gluons). The diagrammatic notation is that of
Ref.~\protect\cite{CSdistfns}}.
\label{evolution}
\end{figure}

The renormalization involves ultraviolet divergent subgraphs, such
as that shown in Fig.~\ref{evolution}(a). Subgraphs with more than two
external parton legs carrying physical polarization, such as that
shown in Fig.~\ref{evolution}(b), do not have an overall divergence.
Thus the divergent subgraphs are the same as for the ordinary parton
distributions. We conclude that the renormalization group equation
for the diffractive parton distributions is
\begin{equation}
\mu { d  \over d\mu}\,
{ d f^{\rm diff}_{a/A}(\xi,\mu;x_\pom,t) \over dx_\pom\,dt}=
\sum_b \int_\xi^1 { dz \over z}\
P_{a/b}(\xi/z,\alpha_s(\mu))\
{ d f^{\rm diff}_{b/A}(z,\mu;x_\pom,t) \over dx_\pom\,dt}
\label{APeqn}
\end{equation}
with the same DGLAP kernel \cite{DGLAP},
$P_{a/b}(\xi/z,\alpha_s(\mu))$, as one uses for the evolution of
ordinary parton distribution functions.

If, following Veneziano and Trentadue, we integrate over $t$, then
the large $t$ integration region introduces new ultraviolet
divergences and the renormalization group equation is modified
\cite{VT}. In this paper, we choose to restrict integrations over
$t$ to the small $t$ region.

\section{Validity of diffractive factorization}
\label{Validity}

In Sec.~\ref{DiffDIS}, we presented the hypothesis of diffractive
factorization for diffractive deeply inelastic scattering, as
represented by Eq.~(\ref{factor}):
\begin{equation}
{d F_2^{\rm diff}(x,Q^2;x_\pom,t)\over dx_\pom\,dt}
\sim \sum_a \int_0^1 d\xi\
{d\, f^{\rm diff}_{a/A}(\xi,\mu;x_\pom,t)\over dx_\pom\,dt}\
\hat F_{2,a}(x/\xi,Q^2;\mu).
\label{factorencore}
\end{equation}
This is an example of the more general hypothesis of factorization for
other kinds of diffractive hard scattering. Another example is
diffractive jet production. Consider, for example, the inclusive cross
section for the production of two jets in a high energy collisions of
two hadrons, $A$ and $B$. (At HERA, this would be $p +\gamma \to
jets$ where hadron $B$ is the the hadronic or ``resolved'' part of the
photon.) Let the initial hadron $A$ have momentum
\begin{equation}
P_{\!A}^\mu = (P_{\!A}^+,P_{\!A}^-,{\bf P}_{\!A})
=  (P_{\!A}^+, {M^2 \over 2P_{\!A}^+},{\bf 0})\,,
\end{equation}
while hadron $B$ enters the scattering with momentum
\begin{equation}
P_{\!B}^\mu = (P_{\!B}^+,P_{\!B}^-,{\bf P}_{\!B})
=  ({M^2 \over 2P_{\!B}^-},P_{\!B}^-, {\bf 0})\,.
\end{equation}
We specify the two jets by variables $E_T$, $X_A$, and $X_B$, given in
terms of the four momenta $P_{\!1}^\mu$ and $P_{\!2}^\mu$ of jets
$1$ and $2$ by
\begin{eqnarray}
E_T &=& (|{\bf P}_{\!1}| + |{\bf P}_{\!2}|)\,,
\nonumber\\
X_A &=& (P_{\!1}^+ + P_{\!2}^+)/P_{\!A}^+\,,
\nonumber\\
X_B &=& (P_{\!1}^- + P_{\!2}^-)/P_{\!B}^-\,.
\end{eqnarray}
If we add the requirement that hadron $A$ emerge scattered with
scattering parameters $(x_\pom,t)$, then we have diffractive jet
production. The corresponding hypothesis of diffractive
factorization for this cross section is
\begin{eqnarray}
\lefteqn{
{d \sigma^{\rm diff}(A + B\to A +{\rm jets} +X)\over
d E_T\, dX_A\, dX_B\, dx_\pom\,dt}
\sim }
\nonumber\\ &&
\sum_{a,b}
\int dx_a\,
{d\,f^{\rm diff}_{a/A}(x_a,\mu;x_\pom,t)\over dx_\pom\,dt}
\int dx_b\, f_{b/B}(x_b,\mu)\
{d \hat\sigma(a + b\to {\rm jets} +X) \over d E_T\, dX_A\, dX_B}\,.
\label{factorjets}
\end{eqnarray}
Of course, Eqs.~(\ref{factorencore}) and  (\ref{factorjets}) are
approximations, as indicated by the $\sim$ signs. We understand the
hypothesis of diffractive factorization to mean that the corrections
to these relations are suppressed by a power of $m/Q$ or
$m/E_T$, where $m$ represents the momentum scale of soft hadronic
interactions and $Q$ or $E_T$ is the scale of the hard interaction.

This general diffractive factorization hypothesis was put forward by
Veneziano and Trentadue in their paper \cite{VT} on fracture
functions. The Ingelman-Schlein model \cite{IS} demands diffractive
factorization plus the Regge structure of the diffractive parton
distributions. Thus, in a strict interpretation, the validity of the
Ingelman-Schlein model logically implies the validity of diffractive
factorization. On the other hand, one might interpret the
Ingelman-Schlein model as being valid if corrections to it, while not
vanishing in the limit of large $Q$, were nevertheless {\it
numerically} small. Thus, for instance, the authors of
Ref.~\cite{BCSS} speculated that the factorization inherent in the
Ingelman-Schlein model was not likely to be exact up to $m/Q$
corrections but might have the same status as Regge factorization,
which has proven to be a useful approximation even if it is not exact.

The hypothesis of diffractive factorization appears to us to be
correct in the case of deeply inelastic scattering. A detailed proof
of this statement is beyond the scope of this paper.  However, we
can briefly sketch how such a proof would go, following the ideas of
Refs.~\cite{CSS,Bodwin}. The singularities of the cut Feynman graphs
for diffractive deeply inelastic scattering are such that the leading
integration regions involve 1) a beam jet in the
direction of the initial hadron $A$ (which includes the final state
diffracted hadron $A^\prime$), 2) a hard interaction, 3) one or more
final state jets that are {\it not} in the direction of hadron $A$,
and 4) possible soft gluons (sometimes with soft quark loops) that may
communicate between the beam jet and the final state jets.
Factorization would be more or less kinematic were it not for the
possibility of the soft gluons linking the beam jet with the final
state jets. One must use gauge invariance to show that the soft
gluons don't really ``see'' the details of the final state jets, so
that the connections to the final state jets can be replaced by
connections to the idealized jet that is embodied in the line
integral of the field operator $A_\mu$ in the definitions of the
diffractive parton distribution functions,
Eqs.~(\ref{quarkdist},\ref{gluedist},\ref{tildepsi},\ref{tildeF}).

\begin{figure}[htb]
\centerline{ \DESepsf(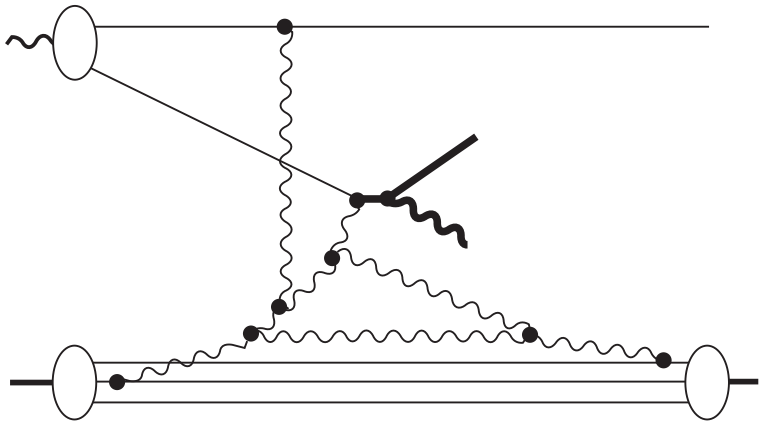 width 10 cm) }
\vskip 0.5 cm
\caption{
A graph for $\gamma + p \to p +{\rm jets} + X$  for which diffractive
factorization fails because of soft color interactions among the
spectator partons.}
\label{diffjetbad}
\end{figure}

What of diffractive factorization for processes with two hadrons in
the initial state, such as $\bar p + p  \to jets$ or $\gamma + p  \to
jets$? Here the proof of {\it ordinary} factorization is much more
delicate. The problem is that low momentum gluons can communicate
between the partons of the two beam jets.  This can happen even
before the hard scattering takes place, as in Fig.~\ref{diffjetbad}.
When one looks for the hard process inclusively, such effects cancel
\cite{CSS,Bodwin}. But the cancellation requires a sum over final
states.  Collins, Frankfurt, and Strikman \cite{CFS} have argued
that the demand that the final state include a diffractively
scattered hadron destroys the factorization.  In Ref.~\cite{BS}, we
looked at this problem in the context of a simple perturbative model.
We found that the diffractive factorization hypothesis is, indeed,
not valid.  New terms proportional to $\delta(1-X_A/x_\pom)$ appear in
Eq.~(\ref{factorjets}). (Presumably in a more general model one will
also have factorization violating terms that are not proportional to
$\delta(1-X_A/x_\pom)$ but are singular as $(1-X_A/x_\pom)\to 0$.)
These new terms have an interesting structure that can be investigated
experimentally at HERA.

\section{ Diffractive parton distributions for $\beta \to 0$}
\label{SmallBeta}

The diffractive parton distribution ${d\,f^{\rm diff}_{a/A}
(\xi,\mu;x_\pom,t)/ dx_\pom\,dt}$ is essentially a long distance
object, which is not amenable to calculation using perturbative
methods. However, Regge phenomenology provides an expectation for
the behavior of ${d\,f^{\rm diff}_{a/A} (\xi,\mu;x_\pom,t)/
dx_\pom\,dt}$ when the parameter
\begin{equation}
\beta = \xi/x_\pom
\end{equation}
is small compared to 1. In the case of the ordinary parton
distributions, this expectation is
\begin{equation}
f_{a/A} (x,\mu) \sim {\rm const.}\times (1/x)^{\tilde\alpha(0)}
\end{equation}
where $\tilde\alpha(0)$ is the pomeron intercept. That is, one
expected
\begin{equation}
F_2(x,Q^2) \sim {\rm const.}\times (1/x)^{\tilde\alpha(0)-1}
\end{equation}
for small $x$. The simplest version of this expectation is that
$\tilde\alpha(0)$ is the same as in soft pomeron physics, $\alpha(0)
-1 \approx 0.08$ \cite{LDsoftpom}. However, this expectation is hard
to reconcile with evolution: if it holds at some rather small value of
$Q^2$, then $F_2$ should be steeper at larger values of $Q^2$
\cite{JCCSmallx}. A steeper dependence was also expected on the basis
of the perturbative version of the pomeron analyzed at leading log
level \cite{BFKL}. Indeed, a steeper dependence, with
$\tilde\alpha(0) -1 \approx 0.4$, is found in experiments at HERA
\cite{HERADIS}.

The analogous expectation \cite{BCSS} for the $\beta$ dependence of
diffractive parton distributions at fixed $x_\pom$ is
\begin{equation}
{d\,f^{\rm diff}_{a/A} (\beta x_\pom,\mu;x_\pom,t)\over dx_\pom\,dt}
\sim {\rm const.}\times  \left(1\over\beta\right)^{\tilde\alpha (0)}.
\end{equation}
In this case, the constant is proportional to the triple pomeron
coupling used in Regge physics. Presumably, $\tilde\alpha(0)$ here
should not be the true pomeron intercept $\approx 1.08$ but
should be the larger value, $\approx 1.4$, found in the HERA
experiments. Then, of course, it is debatable whether the triple
pomeron coupling used here should be the same as that found in soft
Regge physics. Uncertainty over the precise values, however, should
not obscure the simple prediction that the diffractive $F_2$ should
behave at small $\beta$ like
\begin{equation}
{d\,F_2^{\rm diff} (\beta x_\pom,Q^2;x_\pom,t)\over dx_\pom\,dt}
\sim {\rm const.}\times
\left(1\over\beta\right)^{\tilde\alpha (0)-1}.
\end{equation}
with $0 \lesssim \tilde\alpha(0)-1 \lesssim 0.5$. The small $\beta$
behavior of $F_2^{\rm diff}$ is analyzed in Refs.~\cite{smallbeta}.

\section{ Gluon distribution for $\beta \to 1$}
\label{TauGlue}

The diffractive parton distribution ${d\,f^{\rm diff}_{a/A}
(\xi,\mu;x_\pom,t)/ dx_\pom\,dt}$, like the ordinary parton
distribution, is essentially not calculable using perturbative
methods. Recall, however, that it is possible to derive ``constituent
counting rules'' that give predictions for ordinary parton
distributions ${f_{a/A}(x,\mu)}$ in the limit $x \to 1$ for not too
large values of the scale parameter $\mu$ \cite{BrodskyFarrar}. In the
same spirit, we consider in this section and the next the diffractive
parton distributions in the limit $\beta \to 1$ where $\beta =
\xi/x_\pom$. Our analysis is similar to that if Ref.~\cite{BCSS}. The
authors of that paper concluded, with certain caveats, that the
diffractive gluon distribution should behave like $(1-\beta)^1$ as
$\beta \to 1$. Our reanalysis suggests a behavior between
$(1-\beta)^1$ and $(1-\beta)^0$, depending on how certain
nonperturbative issues are resolved. For the diffractive quark
distribution (not treated in  Ref.~\cite{BCSS}), our analysis in
Sec.~\ref{TauQuark} suggests a behavior between $(1-\beta)^2$ and
$(1-\beta)^1$.

This analysis involves both hard and soft subprocesses, put together
in a manner that is not under solid theoretical control. Thus there
is not a clearly correct final answer, as far as we can see. What we
do here is to provide some calculational results that can help
restrict the range of answers and give some basis for the reader's
informed judgment.

Define $\tau = \beta - 1 = \xi/x_\pom - 1$. Then we examine
${d\,f^{\rm diff}_{a/A}((1-\tau)x_\pom,\mu;x_\pom,t)/ dx_\pom\,dt}$
in the limit $\tau \to 0$, after first taking $x_\pom<<1$ so as to
separate out pomeron exchange from other Regge pole exchanges. We
consider the diffractive gluon distribution first ({\it a = g}).
Later, we supply the modifications needed for the diffractive quark
distribution. Throughout this section, except as specifically noted,
we consider Feynman graphs in null plane gauge $A^+ = 0$.

Our analysis is based on a model that consists of a selected set of
Feynman graphs, with the internal loop momenta integrated over a
selected integration region. Within this model there is a hard
subgraph in which all internal propagators are far off-shell and a
soft subgraph in which the propagators are near to being on-shell. We
evaluate the hard subgraph at lowest order in perturbation theory.
We offer a perturbatively based conjecture about the behavior of the
soft subgraph in the Regge limit $x_\pom\to 0$.

The simple graphs considered here do not include the graphs that
contribute to QCD evolution of the parton distributions. Thus we
imagine that the analysis applies to the diffractive parton
distribution at a starting scale $\mu_0$ that is not too large
(say, 2 GeV). Standard parton evolution starting at this scale
will soften the distributions.

\subsection{Decomposition into hard and soft subgraphs}
\label{decompose}

\begin{figure}[htb]
\centerline{ \DESepsf(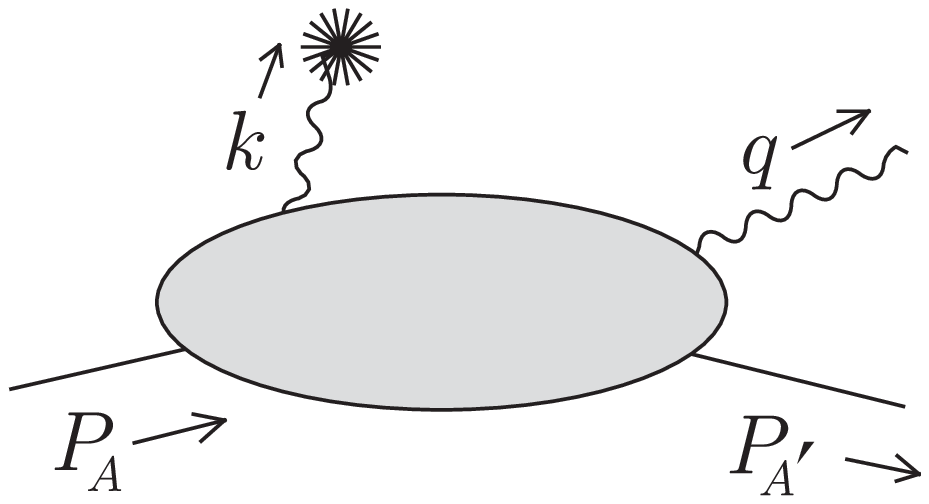 width 8 cm) }
\caption{
Structure of amplitude contributing to the diffractive gluon
distribution. At least one gluon must be emitted into the final
state.}
\vskip 0.5 cm
\label{structurefig}
\end{figure}

According to Eq.~(\ref{fdiff2}), the diffractive gluon distribution
is the square of the matrix element of an operator that destroys a
gluon with longitudinal momentum fraction $(1-\tau)x_\pom$, where the
matrix element is taken between the initial proton state and a final
state that includes the scattered hadron plus anything else. Since a
color octet gluon is destroyed, while the initial and final hadrons
are color singlets, the final state must include at least one gluon.
We consider here the minimal model in which the final state includes
precisely one gluon. Then this gluon carries a very small momentum
fraction $\tau x_\pom$. Call the momentum of the final state gluon
$q^\mu$, as depicted in Fig.~\ref{structurefig}. Since this gluon is
on-shell, we have
\begin{equation}
q^\mu = (\tau x_\pom P_{\!A}^+, {{{\bf q}^2} \over {2\tau x_\pom
P_{\!A}^+}},
{\bf q}).
\end{equation}

Before proceeding, we pause for a technical point. In general in this
section we use the null plane gauge $A^+=0$, but for the final state
gluon this is not convenient. The polarization vectors
$\epsilon^\mu(q,j)$ for transverse polarization in the $j$ direction
have minus components that grow like $1/\tau$ in the limit
$\tau \to 0$:
\begin{equation}
\epsilon^-(q,j) = { q^j \over q^+}
= { q^j \over x_\pom\tau P_{\!A}^+}.
\end{equation}
We avoid this singular behavior by changing the polarization vectors
for this gluon to $A^- = 0$ gauge. The difference $\Delta
\epsilon^\mu$ between the old polarization vector and the new is
proportional to $q^\mu$, so that changing polarization vectors has no
effect after we sum over a gauge invariant set of graphs. In $A^- =
0$ gauge, $\epsilon^-(q,j) = 0$, $\epsilon^i(q,j)= \delta^{ij}$ and
\begin{equation}
\epsilon^+(q,j) = { q^j \over q^-}
= x_\pom\tau P_{\!A}^+\,{ q^j \over {\bf q}^2}.
\label{Aminusgauge}
\end{equation}
Thus $\epsilon^\mu(q,j)$ is predominantly transverse, with a
plus component that vanishes as $\tau \to 0$.

Some vocabulary will be helpful for discussing the physics of time
and momentum scales in this problem.  There are three relevant
longitudinal momentum scales. We call partons with $p^+ \sim
P_{\!A}^+$ fast partons. For instance, the valence quarks in a hadron
are typically fast partons. We call partons with  $p^+ \sim x_\pom
P_{\!A}^+$ slow partons. Finally, we call partons with $p^+ \sim \tau
x_\pom P_{\!A}^+$ very slow partons. The final state gluon is such a
parton.

The gluon cloud surrounding any hadron contains gluons at any
momentum fraction. Consider a gluon with momentum fraction $x$
and transverse momentum of order $m$, where $m \approx 0.3\
{\rm GeV}$ gives the scale of a typical hadronic mass or transverse
momentum. The contribution of such a gluon to the null plane energy
$P^-$ of an intermediate state is $p^- = {\bf p}^2/2p^+$, which is
the minus momentum of a free gluon with transverse momentum ${\bf p}$
and plus momentum $p^+$. This ``kinetic'' minus momentum is of order
$m^2 /(xP_{\!A}^+)$. Thus a gluon of the type that we are calling a slow
gluon has a kinetic null plane energy $p^- \sim m^2 /(x_\pom P_{\!A}^+)$
and survives for a typical null plane time $\Delta z^+ \sim x
P_{\!A}^+/m^2$.

Notice that the final state gluon has a large minus momentum, $q^-
={{{\bf q}^2} / ({2\tau x_\pom P_{\!A}^+})} $, at least as long
as its transverse momentum $\bf q$ is not too small. Now $\bf q$ is
not observed; we are to square the matrix element and integrate over
$\bf q$. We cannot say anything about the region of very small $\bf
q$, but we can analyze the contribution to the integral from the
region $\cal R$ defined by
\begin{equation}
{\bf q}^2 \gg \tau m^2.
\label{calR}
\end{equation}
We will consider the contribution to $df^{\rm diff}/ dx_\pom\,dt$
from the region $\cal R$ with the hope that the contribution from the
complementary region ${\bf q}^2\lesssim \tau m^2$, is not large enough
to overwhelm --- or, worse, to cancel --- the contribution from
region $\cal R$.

We will also evaluate the contribution from the smaller integration
region in which the transverse momentum ${\bf q}$ is large: ${\bf q}^2
\gg m^2$.  Since this region is described by rather standard short
distance dynamics, its contribution should provide a lower bound on
the true result. (Again, we assume that this contribution is not
canceled by some long distance contribution.)

For ${\bf q} \in {\cal R}$ the minus momentum $q^-={{{\bf q}^2} /
({2\tau x_\pom P_{\!A}^+})} $ of the final state gluon is large
compared to the kinetic minus momentum $p^- \sim m^2/P_{\!A}^+$ of a
typical fast parton and is also large compared to the kinetic minus
momentum $p^- \sim m^2/(x_\pom P_{\!A}^+)$ of a typical slow parton.
This large minus momentum flows through the graph and is carried out
of the graph by the detected gluon. Thus the parton detection with
$\tau \to 0$ creates a hard process that happens on a null plane time
scale $\Delta x^+$ that is short compared to the typical time scale
for interactions within the proton or its cloud of slow gluons. We
base our analysis on this observation, using low order perturbation
theory for the nearly instantaneous interaction that probes the
parton distribution.

\begin{figure}[htb]
\centerline{ \DESepsf(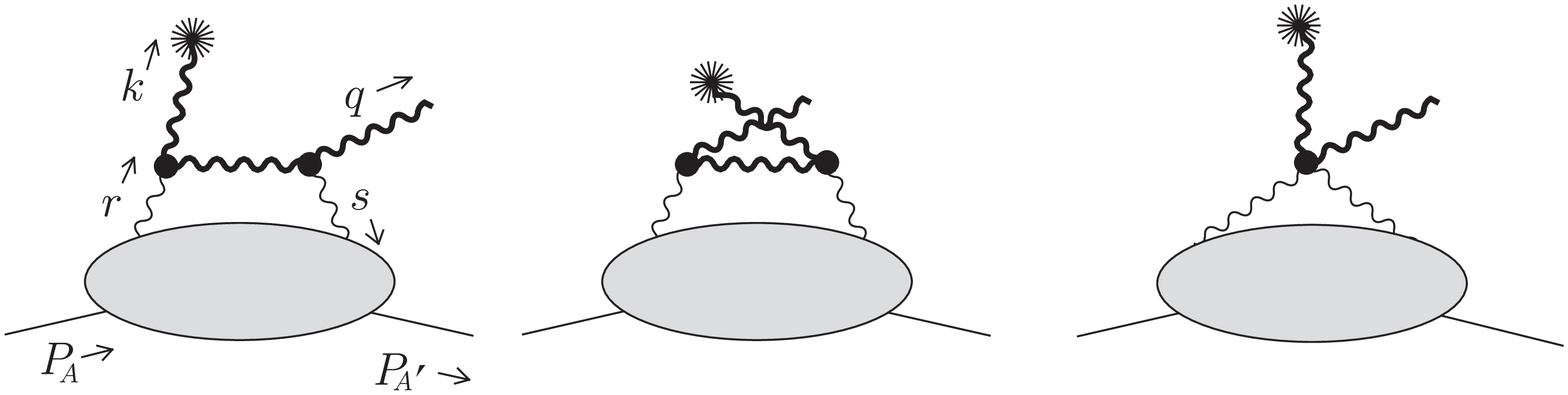 width 15 cm) }
\caption{
Graphs with an order $g^2$ hard subgraph.}
\vskip 0.5 cm
\label{hardfig}
\end{figure}

Three graphs that involve the lowest order hard interaction are
shown in Fig.~\ref{hardfig}. In these graphs, the large minus momentum
$q^-$ flows through only one or two propagators, as indicated by the
heavy lines. These propagators are far off shell. We refer to this
part of the graph as the hard subgraph. The rest of the graph is the
soft subgraph.

\begin{figure}[htb]
\centerline{ \DESepsf(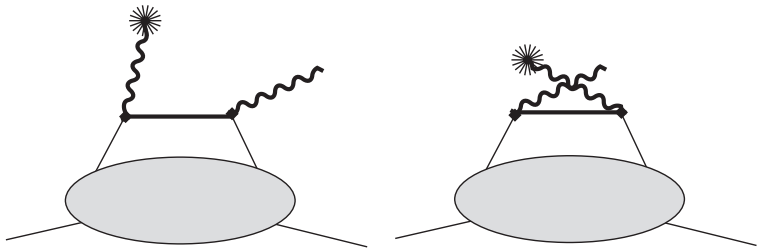 width 10 cm) }
\vskip 0.5 cm
\caption{
More graphs with an order $g^2$ hard subgraph. These graphs involve
quark exchange from the soft subgraph and are not considered in
this paper.}
\label{slowquarkfig}
\end{figure}

There are also two graphs of the same order that involve connecting
the $q^\mu$ and $k^\mu$ gluons to a quark line, as depicted in
Fig.~\ref{slowquarkfig}. These graphs require a separate discussion,
which we omit in this paper on the grounds that low momentum gluons
are likely to be dominant in the pomeron compared to low momentum
quarks.

In Fig.~\ref{hardfig}, we denote by $r^\mu$ and $s^\mu$,
respectively, the momenta of the gluon leaving the soft subgraph and
entering it after the hard interaction. Let the momentum fraction of
the first gluon be $r^+/P_{\!A}^+ = (1+\sigma) x_\pom$. Then momentum
conservation fixes $s^+/P^+ = \sigma x_\pom$.

We integrate over $\sigma$. It is convenient to distinguish between
the two identical gluons by requiring that $r^+ > - s^+$. That is,
$\sigma > - 1/2$. In principle, the $\sigma$ integration runs to
$\sigma = \infty$, but the reader can verify that the region
$\sigma \gg 1$ is not important. We take it as a working assumption,
to be verified later, that the region $|\sigma| \ll 1$ is also not
important. Thus the important integration region is $\sigma \sim 1$.
That is, the gluons that couple the hard subgraph to the soft
subgraph are typical ``slow'' gluons.

We integrate over the transverse momentum ${\bf r}$, setting ${\bf s}=
{\bf r} + {\bf l}$. We also integrate over $r^-$, setting $s^- = r^-
- M_{\!A}^2/(2P_{\!A}^+) +( M_{\!A}^2 +({\bf P}_{\!A^\prime})^2)/(2
(1-x_\pom) P_{\!A}^+)$. We suppose that the hadron wave functions fix
${\bf r}$ and $r^-$ to be no larger than the ordinary size for slow
gluons, ${\bf r}^2 \sim m^2$ and $r^- \lesssim m^2/(x_\pom
P_{\!A}^+)$. (Contributions in which $r^\mu$ is part of a hard
virtual loop are more properly considered to be part of a higher
order correction to the hard subgraph.)

Taking the limit $\tau \to 0$, we find that if we consider the hard
subgraph to be a function of $r^\mu$, $s^\mu$, and $q^\mu$, then it
is independent of $r^-$, $s^-$, ${\bf r}$ and ${\bf s}$ and is also
also independent of $x_\pom$. In addition, the dominant
polarizations for the gluons entering the hard subgraph from the
soft subgraph are transverse; in a shorthand notation,
\begin{equation}
\sum_{\mu,\nu = \{+,-,1,2\}}
[{\rm Hard}]^{\mu\nu}[{\rm Soft}]_{\mu\nu}
\approx
\sum_{i,j = \{1,2\}}[{\rm Hard}]^{ij}[{\rm Soft}]_{ij}.
\label{HardSoftindices}
\end{equation}
That the hard subgraph is independent of $r^-$  and $s^-$ is
not surprising, since $r^- \ll q^-$ and $s^- \ll q^-$. That it is
independent of $x_\pom$ follows simply from its invariance under
boosts in the longitudinal direction. It is not, however, obvious
that the hard subgraph is independent of ${\bf r}$ and ${\bf s}$ and
that only transverse polarizations are important. These results
follow from an argument that we describe in
Appendix~\ref{Appendix:HardStructure}. The picture that emerges is
one in which the soft gluon cloud surrounding the hadron is probed by
an interaction that is effectively local in null plane time, $x^+$
and in transverse position $\bf x$.

\subsection{Structure of the diffractive gluon distribution}
\label{structure}

\begin{figure}[htb]
\centerline{ \DESepsf(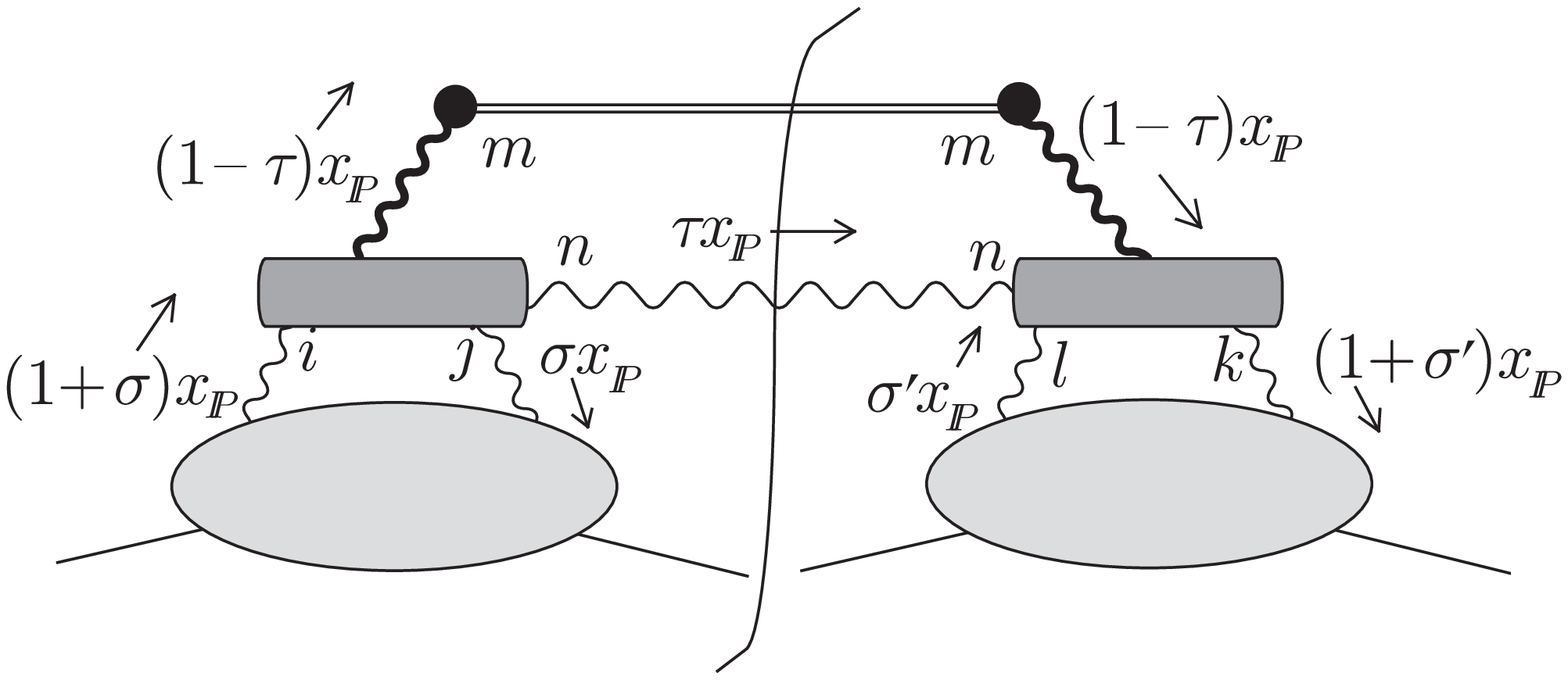 width 15 cm) }
\vskip 0.5 cm
\caption{
Factored structure of the diffractive gluon distribution function
for $\tau \to 1$.}
\label{diffgluefig}
\end{figure}

We take advantage of these results by setting $r^- = s^- = 0$ and
${\bf r} = {\bf s} = {\bf 0}$ in the hard subgraph and restricting
the polarization sum to transverse polarizations. Then the
diffractive distribution function depends on the integral over $r^-$
and ${\bf r}$ of the soft subgraph. We write this structure as
depicted in Fig.~\ref{diffgluefig},
\begin{eqnarray}
{{df_{g/A}^{\rm diff}} \over {dx_\pom\,dt}} &=&
{1 \over {64\pi^3\tau}}
\,{ 1 \over 2}\sum_{s_{\!A},s_{\!A^\prime}}\,
\sum_{ijkl}
\int_{-1/2}^\infty d\sigma^\prime
A_{kl}^*(\sigma^\prime,x_\pom,t;s_{\!A},s_{\!A^\prime})
\int_{-1/2}^\infty d\sigma
A_{ij}(\sigma,x_\pom,t;s_{\!A},s_{\!A^\prime})
\nonumber\\
&&\times
\sum_{mnab}
\int_{\cal R} {{d^2{\bf q}} \over {(2\pi)^2}}
{{{\cal M}_{ab}^{klmn}({\bf q},\sigma^\prime,\tau)^*\
{\cal M}_{ab}^{ijmn}({\bf q},\sigma,\tau)}
\over {(k^2 - i\epsilon)\ (k^2 + i\epsilon)}}.
\label{diffint}
\end{eqnarray}
Here the first line represents the soft subgraphs while the second line
represents the integral of the hard subgraphs. We discuss these factors
below.

The function $A_{ij}(\sigma,x_\pom,t;s_{\!A},s_{\!A^\prime})$ is the
amplitude for the proton to emit a gluon with transverse polarization
$i$ and momentum fraction $(1+\sigma)x_\pom$ then absorb a gluon with
transverse polarization $j$ and momentum fraction $\sigma x_\pom$.
\begin{eqnarray}
A^{ij}(\sigma,x_\pom,t;s_{\!A},s_{\!A^\prime}) &=&  x_\pom P_{\!A}^+
\sum_a\int dy^-
e^{-i\sigma x_\pom P_{\!A}^+y^-}
\nonumber\\
&&\quad \times
\langle P_{\!A^\prime},s_{\!A^\prime}|{\rm T}\left\{A_a^{j}(0,y^-,0_T)\,
A_a^{i}(0)\right\}|P_{\!A},s_{\!A} \rangle.
\label{Aij}
\end{eqnarray}
There is a summation over the color $a$ of the gluon operators; in
$\cal M$ we average over these colors. The factor $x_\pom P_{\!A}^+$
arises from changing the integration in Eq.~(\ref{diffint}) from
$\int d r^+$ to $\int d\sigma$.

We now turn to the hard interaction. We begin by considering the probed
gluon, which carries momentum $k^\mu$. According to
Eqs.~(\ref{fdifftoG}) and (\ref{fdiff2}), the operator that probes the
gluon distribution is $\tilde F_a^{+m}(y)$.  (Only $\nu \in \{1,2\}$
contributes in Eq.~(\ref{fdiff2}).)  In $A^+=0$ gauge, this is simply
$\partial^+ A_a^m(y)$. (We have changed the polarization vector
$\epsilon (q,n)$ for the final state gluon from $\epsilon^+ = 0$, so
this gluon could couple to the $A^+ A^m$ operator in $\tilde
F_a^{+m}(y)$. However, this graph is not allowed when the two exchanged
gluons are in a color singlet state.) The $\partial^+$ becomes a $k^+ =
(1-\tau) x_\pom P_{\!A}^+ \sim x_\pom P_{\!A}^+$ that we absorb into the
normalization. This leaves a propagator for the probed gluon,
\begin{equation}
{ -i D^{m\mu}(k) \over k^2 + i\epsilon}
={ -i \over k^2 + i\epsilon}
\sum_{j^\prime}\epsilon^m(k,j^\prime)\,\epsilon^\mu(k,j^\prime)
 = {-i \over k^2 + i\epsilon}\,\epsilon^\mu(k,m).
\end{equation}
Thus each probe operator gives a $1/k^2$ factor and leaves the
amputated hard interaction graph dotted into a polarization vector
for a gluon of momentum $k^\mu$, polarization $m$, in $A^+=0$ gauge.

We call the amputated hard interaction graph ${\cal M}_{ab}^{ijmn}
({\bf q},\sigma,\tau)$.  The indices $m$ and $n$ are the transverse
polarizations of the probed and final state gluons respectively; $a$
and $b$ are their colors. Recall that the final state gluon
polarization is represented by a polarization vector
$\epsilon^\mu(q,n)$ is, by our convention, in $A^- = 0$ gauge, as in
Eq.~(\ref{Aminusgauge}). In $\cal M$, the exchanged gluons are
approximated as having zero transverse and minus momenta; one
integrates over these momenta in the definition of $A$.  The exchanged
gluons have transverse polarizations $i$ and $j$ respectively.

In the $\tau \to 0$ limit, the factor $k^2$ in the denominator is
\begin{equation}
k^2 \sim -{\bf q}^2/\tau.
\end{equation}
Thus the probed gluon is far off shell, $k^2 \gg m^2$, as long as $\bf
q$ is in the integration region $\cal R$, ${\bf q}^2 \gg \tau m^2$.
Similarly, the virtual lines internal to $\cal M$ are far off shell and
have a simple form when ${\bf q} \in {\cal R}$. This is so even though
the region $\cal R$ includes small transverse momenta, ${\bf q}^2 \ll
m^2$. Indeed, the small transverse momenta near the boundary of $\cal
R$ dominate the integral in Eq.~(\ref{diffint}), as we will see.

Next, we examine the soft subgraphs and then the hard subgraphs in
some detail.

\subsection{The soft subgraph}
\label{soft}

In this subsection, we investigate the $x_\pom$ dependence of
$A^{ij}(\sigma,x_\pom,t;s_{\!A},s_{\!A^\prime})$ in the limit $x_\pom
\to 0$. Comparing the expected Regge form (\ref{pomdist}) of
${{df_{g/A}^{\rm diff}} / {dx_\pom\,dt}}$ with Eq.~(\ref{diffint}),
we see that the expected Regge form of $A^{ij}$ is
\begin{equation}
A^{ij}(\sigma,x_\pom,t;s_{\!A},s_{\!A^\prime}) \propto
x_\pom^{-\alpha(t)}.
\label{zdependenceofA}
\end{equation}
The question is, what is the pomeron trajectory $\alpha(t)$ within
the context of the analysis used in this paper?

The operators Eq.~(\ref{Aij}) are in $A^+ = 0$ gauge, but the
amplitude can be put in a gauge invariant form by reexpressing it in
terms of the gluon field operators $\tilde F^{\mu\nu}$ defined in
Eq.~(\ref{tildeF}). We simply replace
\begin{equation}
A^j \to
{ s^+ A^j \over s^+ +i\epsilon}
\to -i{ \partial^+ A^j \over s^+ +i\epsilon}
\to -i{ \tilde F^{+j} \over s^+ +i\epsilon}
\label{Ajreplacement}
\end{equation}
and make a similar replacement for $A^i$. The $i\epsilon$ choice
here is of some significance. We will discuss it in the following
subsection. The resulting form for $A^{ij}$ is
\begin{eqnarray}
A^{ij}(\sigma,x_\pom,t;s_{\!A},s_{\!A^\prime}) &=&
{1 \over { x_\pom(\sigma+i\epsilon)(1+\sigma) P_{\!A}^+ }}
\sum_a\int dy^-
e^{-i\sigma x_\pom P_{\!A}^+y^-}
\nonumber\\
&&\quad \times
\langle P_{\!A^\prime},s_{\!A^\prime}|{\rm T}
\left\{\tilde{F}_a^{+j}(0,y^-,0_T)
\tilde{F}_a^{+i}(0)\right\}|P_{\!A},s_{\!A} \rangle.
\label{Aijmod}
\end{eqnarray}
Since the propagators in the soft subgraph are $\it not$ far
off-shell, one is not really justified to use perturbation theory
to investigate $A^{ij}(\sigma,x_\pom,t;s_{\!A},s_{\!A^\prime})$.
Nevertheless, perturbation theory is suggestive. Consider
Eq.~(\ref{Aijmod}) using Feynman gauge. Suppose that the gluon
annihilated by the operator $\tilde{F}_a^{+i}$ connects to an on-shell
quark carrying no transverse momentum and plus momentum $p^+ =
\lambda P_{\!A}^+$, with $\lambda$ of order 1. That is, the gluon
couples to a ``fast'' quark. The relevant factor is
\begin{equation}
{- r^i \over r^2+i\epsilon}\
\bar {\cal U}(\lambda^\prime P_{\!A}^\mu,s)\,
igt_c\gamma^+\,
{\cal U}(\lambda P_{\!A}^\mu,s)\,.
\end{equation}
where $\lambda^\prime = \lambda - x_\pom (1+\sigma)$.  In the limit
$x_\pom \to 0$, this is independent of $x_\pom$. Similarly, the
coupling of the operator $\tilde{F}_a^{+j}$ to a fast quark gives no
$x_\pom$ dependence. Thus perturbation theory suggests that the
operator matrix element in Eq.~(\ref{Aijmod}) is independent of
$x_\pom$ for small $x_\pom$. Considering that there is a factor
$1/x_\pom$ in Eq.~(\ref{Aijmod}), we find that the pomeron trajectory
appearing in Eq.~(\ref{zdependenceofA}) is $\alpha(t) = 1$ in this
most naive analysis of the soft subgraph. This is close to the
pomeron trajectory observed in nature, but of course this most naive
analysis is too naive, and we expect that soft interactions among the
gluons modify the result. What we see here is that the picture of the
pomeron as two gluon exchange, which often gives results that are
surprisingly good considering the simplicity of the picture
\cite{LowNussinov}, works rather well also in this context.

\subsection{Small $\sigma$ singularity}
\label{singularity}

The analysis of the preceding subsection has revealed a $1/\sigma$
singularity within the integration domain $-1/2 < \sigma < \infty$ of
the momentum fraction $\sigma$. What is the nature of this
singularity? If we stick to $A^+ = 0$ gauge, one can check that it
arises from the $1/s^+$ singularity in the gluon propagator,
\begin{equation}
{i \over s^2 + i\epsilon}
\left [
-g^{\mu\nu}
+{g^{\mu +}s^\nu +s^\mu g^{\nu +}\over s^+}
\right].
\end{equation}
This singularity is usually interpreted with a principle value
prescription, but there is no compelling reason for this choice.
In fact, $A^+ = 0$ gauge is not an effective tool for
examining the nature of the $1/\sigma$ singularity, since in this
gauge the singularity is a gauge artifact. Thus we choose to
examine this question in Feynman gauge, in the style of
Ref.~\cite{CSS}. (Arguably, it would be better to do the whole problem
in Feynman gauge, but this leads to its own complications.)

\begin{figure}[htb]
\centerline{ \DESepsf(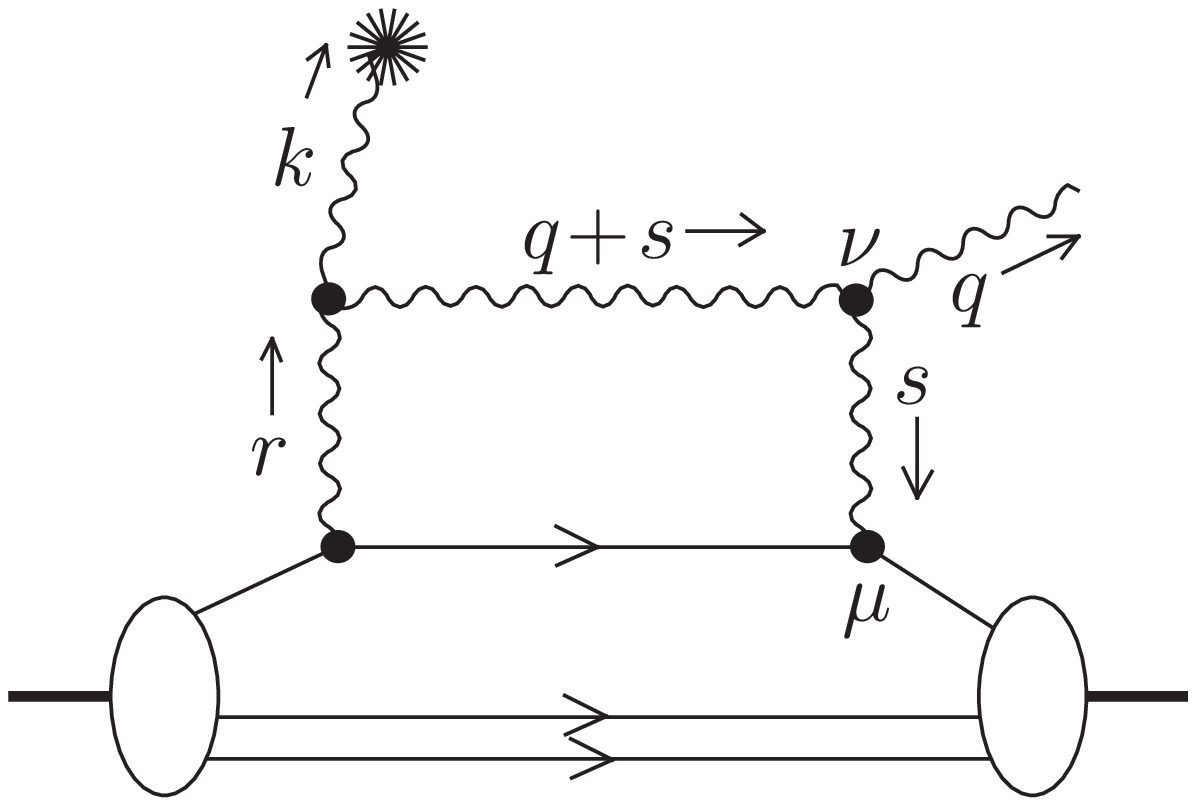 width 10 cm) }
\vskip 0.5 cm
\caption{
Graph with a singularity near $s^+ = 0$.}
\label{singularityfig}
\end{figure}

We consider the graph shown in Fig.~\ref{singularityfig} in Feynman
gauge but with a transverse polarization chosen for the gluon carrying
momentum $r^\mu$. It is helpful to choose a frame in which $x_\pom
P_{\!A}^+ \sim m$. Then
\begin{eqnarray}
P_{\!A^\prime}^{+}= (1-x_\pom) P_{\!A}^+\gg m
&&\quad
P_{\!A^\prime}^{-} = {\bf P}_{\!A^\prime}^2 /[2 (1-x_\pom) P_{\!A}^+]\ll m
\nonumber\\
q^+ = \tau x_\pom P_{\!A}^+ \ll m
&&\quad
q^- = {\bf q}^2/[2 \tau x_\pom P_{\!A}^+] \gg m.
\end{eqnarray}
The graph contains the structure
\begin{equation}
J^\mu\, {-i g_{\mu\nu} \over s^2 + i\epsilon}\,
{N^\nu\over (q+s)^2 + i\epsilon}.
\end{equation}
The denominator $(q + s)^2$ has the form
\begin{equation}
(q+s)^2 + i\epsilon
= 2 q\cdot s + s^2 + i\epsilon
= 2 (q^- + s^-) s^+ - {\bf q}\cdot {\bf s}
+ 2q^+s^- - {\bf s}^2 + i\epsilon
\end{equation}
Now $q^-$ is large and positive while $s^-$ is small in the dominant
integration region. Thus the denominator has the approximate form
\begin{equation}
{1 \over 2q^-\,[s^+ + \cdots + i\epsilon]}
\label{sdenom}
\end{equation}
where the dots indicate small terms. This denominator contains the
only important dependence on $s^+$ as long as $|\sigma| \lesssim 1$
in $s^+ = \sigma x_\pom P_{\!A}^+$. Notice that there is a singularity
very near to $s^+ = 0$, but that, since there are no other
singularities nearby, we can deform the integration contour away from
this singularity. Thus we deform the $s^+$ contour into the upper
half complex $s^+$ plane. On the deformed contour, we have
$s^+ \sim m$ or $\sigma \sim 1$.

Now we examine the numerators. The largest component of the quark
current $J^\mu$ is $J^+$, while the largest component of the current
$N^\nu$ of the final state gluon is $N^-$.  Thus $J_\mu N^\mu \sim
J^+ N^-$. Thus the dominant term in $J_\mu N^\mu$ is obtained by
replacing
\begin{equation}
J_\mu N^\mu \to {J^+ \over s^+}\ s_\mu N^\mu.
\label{replacement0}
\end{equation}
The factor $s_\mu N^\mu$ is approximately $s^+ N^-$ as long as $s^+$
is not small, and we know that $s^+$ {\it is} not small since we have
deformed the integration contour so that it does not approach $s^+ =
0$.

The next step is to restore the $s^+$ integration contour to the
real axis, taking care not to cross any singularities. This means
that we should move the $1/s^+$ singularity in
Eq.~(\ref{replacement0}) infinitesimally into the lower half $s^+$
plane, so that our replacement becomes
\begin{equation}
J_\mu N^\mu \to {J^+ \over s^+ +i\epsilon}\ s_\mu N^\mu.
\label{replacement}
\end{equation}
This is in keeping with the usual notation in which integration
contours are along the real axis, with poles infinitesimally displaced
from the integration contour.

The replacement (\ref{replacement}) gives the dominant contribution
to our graph. However, if we attach the gluon carrying momentum
$s^\mu$ everywhere in the hard subgraph and sum the leading terms
obtained with this replacement, we will get zero because of the Ward
identities obeyed by the hard graph and because the two gluons
carrying momenta $r^\mu$ and $s^\mu$ together form a color singlet.
(The relevant identities are discussed in the appendix of
Ref.~\cite{CSS}).

We have thus encountered the {\it b\^ete noir} of Feynman gauge: the
leading contributions graph by graph come from unphysical
polarizations that cancel when one sums over graphs. What we need is
the subleading contribution. That is easy. We replace
\begin{equation}
J_\mu N^\mu = {J^+ \over s^+ +i\epsilon}\ s_\mu N^\mu
+ \tilde J_\mu N^\mu.
\label{rereplacement}
\end{equation}
where
\begin{equation}
\tilde J^\mu =
{s^+ J^\mu - J^+ s^\mu \over s^+ +i\epsilon}.
\end{equation}
Now we throw away the first term in (\ref{rereplacement}), since it
will cancel, and keep the remainder. This gives the structure in
Eq.~(\ref{Ajreplacement}), but now with a prescription for the
$1/s^+$ singularity determined on physical grounds.

\subsection{The hard subgraph}
\label{hard}

We now turn to the hard interaction function $\cal M$.
It is a simple matter to evaluate this function. We find
\begin{eqnarray}
\lefteqn{
{\cal M}_{ab}^{ijmn}({\bf q},\sigma,\tau)=}
\nonumber\\
&&{3ig^2 \over 4(\sigma+i\epsilon ) (1+\sigma)}\,\delta_{ab}
\biggl[
\sigma(1+\sigma)\ \delta^{ij}\,\delta^{mn}
- \sigma\ \delta^{in}\,\delta^{jm}
+ (1+\sigma)\ \delta^{im}\,\delta^{jn}
\biggr].
\label{calMresult}
\end{eqnarray}
Thus ${\cal M}\propto \tau^0$ as $\tau \to 0$. The $i\epsilon$
prescription in the $1/\sigma$ factor arises from Eq.~(\ref{sdenom}).
It matches the $i\epsilon$ prescription in the soft function $A_{ij}$,
Eq.~(\ref{Aijmod}), so that, after a contour deformation, $\sigma$ is
never much smaller than 1.

Inserting Eq.~(\ref{calMresult}) and $k^2 \sim -{\bf q}^2/\tau$ into
Eq.~(\ref{diffint}), we obtain
\begin{eqnarray}
{{df_{g/A}^{\rm diff}} \over {dx_\pom\,dt}} &=&
{1 \over {64\pi^3\tau}}
\,{ 1 \over 2}\sum_{s_{\!A},s_{\!A^\prime}}\,
\sum_{ijkl}
\nonumber\\
&&\times \int_{-1/2}^\infty d\sigma^\prime
A_{kl}^*(\sigma^\prime,x_\pom,t;s_{\!A},s_{\!A^\prime})\,
\int_{-1/2}^\infty d\sigma
A_{ij}(\sigma,x_\pom,t;s_{\!A},s_{\!A^\prime})\
{\cal H}^{klij}(\sigma,\sigma^\prime)
\nonumber\\
&&\times
\int {{d^2{\bf q}} \over {(2\pi)^2}}
{ \theta({\bf q}^2 > \tau M^2) \over ({\bf q}^2/\tau )^2 }.
\label{diffint2}
\end{eqnarray}
Here ${\cal H}^{klij}(\sigma,\sigma^\prime)$ is a rational function of
$\sigma$ and $\sigma^\prime$ that is not of particular interest.
Recall that we integrate $\bf q$ over the region $\cal R$ defined by
${\bf q}^2 \gg \tau m^2$ (Eq.~(\ref{calR})), where $m$ is a typical
hadronic mass or transverse momentum. Here we make the prescription
more precise by integrating over ${\bf q}^2 > \tau M^2$ where $M$ is
any fixed mass such that $M\gg m$.

\subsection{Result}
\label{result_gluon}
Performing the integration in Eq.~(\ref{diffint2}) gives for the $\tau \to
0$ limit of the diffractive gluon distribution
\begin{eqnarray}
{{df_{g/A}^{\rm diff}} \over {dx_\pom\,dt}} &\sim&
{1 \over {256\pi^4\,M^2}}
\,{ 1 \over 2}\sum_{s_{\!A},s_{\!A^\prime}}\,
\sum_{ijkl}
\nonumber\\
&&\times
\int_{-1/2}^\infty d\sigma^\prime
A_{kl}^*(\sigma^\prime,x_\pom,t;s_{\!A},s_{\!A^\prime})\,
\int_{-1/2}^\infty d\sigma
A_{ij}(\sigma,x_\pom,t;s_{\!A},s_{\!A^\prime})\
{\cal H}^{klij}(\sigma,\sigma^\prime).
\label{net_gluon}
\end{eqnarray}
Note that ${df_{g/A}^{\rm diff}} / {dx_\pom\,dt}$ is independent of
$\tau$ as $\tau \to 0$:
\begin{equation}
{{df_{g/A}^{\rm diff}}(\beta x_\pom,\mu;x_\pom,t)
 \over {dx_\pom\,dt}}
 \sim (1-\beta)^0
  \quad\quad {\rm as}\ \beta \to 1\ {\rm with}\ x_\pom = const.
\label{ginpomresult}
\end{equation}
This may seem surprising, since the distribution of gluons in a typical
hadron behaves like $f_{g/A}(x;\mu_0) \sim {\rm const.} \times
(1-x)^p$ with a rather high power $p \approx 5$.

The leading behavior comes from the lower endpoint of the integration
in Eq.~(\ref{diffint2}). Thus it is sensitive to the cutoff chosen.
Recall that we took ${\bf q}^2 > \tau M^2$ because as long as ${\bf
q}^2/\tau  \gg m^2$, the internal lines in the hard subdiagram are far
off-shell. This appears to be the natural cutoff. As soon as the
internal lines of the subdiagram through which $q^\mu$ flows are {\it
not} far off-shell, the the gluon that goes into the final state and
the probed gluon can attach at different space-time points in the
gluon cloud of the hadron. Then there is the opportunity for
cancellation, as both gluons sample the color charge of the hadron as
a whole and find that the hadron as a whole is a color singlet. It is
difficult to check this conjecture directly in a realistic model of
nonperturbative hadron structure. However we have checked in a very
simple model where all the graphs can be included exactly. This model
is too simple to have the correct pomeron behavior, but we find that
it does have $\tau^0$ behavior in the $\tau \to 0$ limit. The model is
described in Appendix \ref{Appendix:Model}.

One might reasonably conjecture that a larger cutoff would be imposed
by nonperturbative physics in a more realistic model.  For instance, if
the gluon emitted into the final state effectively had a substantial
mass $m_g$, then $q^-$ would be $({\bf q}^2 + m_g^2)/(2x_\pom\tau
P_{\!A}^+)$. This would induce an effective cutoff ${\bf q}^2 > m_g^2$ in
Eq.~(\ref{diffint2}). Then we would have obtained
${{df_{g/A}^{\rm diff}}(\beta x_\pom,\mu;x_\pom,t)/ {dx_\pom\,dt}}
\propto (1-\beta)^p$ with $p = 1$. This corresponds to the style of
analysis of the constituent counting rules and gives the result
$(1-\beta)^1$ found in Ref.~\cite{BCSS}. Presumably, the contribution
from ${\bf q}^2 \gg m^2$ must be present and is not likely to be
canceled, so that ${{df_{g/A}^{\rm diff}}(\beta
x_\pom,\mu;x_\pom,t)/ {dx_\pom\,dt}}$ should not be smaller than
$(1-\beta)^1$ at large $\beta$. That is, the power $p$ should not be
larger than 1.

We conclude that if the diffractive gluon distribution is
parameterized as
\begin{equation}
{{df_{g/A}^{\rm diff}}(\beta x_\pom,\mu;x_\pom,t) \over dx_\pom\,dt}
 \propto (1-\beta)^p
\end{equation}
for $\beta \to 1$ at moderate values of the scale $\mu$, say 2
GeV, then
\begin{equation}
0 \lesssim p \lesssim 1.
\end{equation}
The choice $p\approx 0$ corresponds to an effectively massless final
state gluon, while $p\approx 1$ corresponds to an effective gluon
mass.

\section{Quark distribution for $\beta \to 1$}
\label{TauQuark}

In analogy with the gluon case, we write
\begin{eqnarray}
df_{q/A}^{\rm diff} \over {dx_\pom\,dt} &=&
{1 \over {64\pi^3\tau}}
\,{ 1 \over 2}\sum_{s_{\!A},s_{\!A^\prime}}\,
\int_{-1/2}^\infty d\sigma
\int {{d^2{\bf s}} \over {(2\pi)^2}}\
\int_{-1/2}^\infty d\sigma^\prime
\int {{d^2{\bf s}^\prime} \over {(2\pi)^2}}\
\nonumber\\
&&\times
S_{\rho\sigma}^*(\sigma^\prime,{\bf s}^\prime
;x_\pom,t;s_{\!A},s_{\!A^\prime}) S_{\mu\nu}(\sigma,{\bf
s};x_\pom,t;s_{\!A},s_{\!A^\prime})
\nonumber\\
&&\times
\sum_{s_q\,s_k}\sum_{IJ}
\int {{d^2{\bf q}} \over {(2\pi)^2}}\
\nonumber\\
&&\times
{{{\cal M}_{IJ}^{\rho\sigma}
({\bf q},{\bf k},{\bf s}^\prime,\sigma,\tau;s_q,s_k)^*\
{\cal M}_{IJ}^{\mu\nu}({\bf q},{\bf k},{\bf s},\sigma,\tau;s_q,s_k)}
\over {(k^2 - i\epsilon)\ (k^2 + i\epsilon)}}.
\label{diffintquark}
\end{eqnarray}
Here the soft function $S_{\mu\nu}(\sigma,{\bf s};x_\pom,
t;s_{\!A},s_{\!A^\prime})$ is the amplitude for the proton to emit a
gluon with polarization $\mu$, momentum fraction $(1+\sigma) x_\pom$
and transverse momentum ${\bf r} = {\bf s} - {\bf P}_{\!A^\prime}$,
then absorb a gluon with transverse polarization $\nu$ and momentum
fraction $\sigma x_\pom$ and transverse momentum ${\bf s}$:
\begin{eqnarray}
S_{\mu\nu}(\sigma,{\bf s};x_\pom,t;s_{\!A},s_{\!A^\prime})
&=&  x_\pom P_{\!A}^+
\sum_a\int dy^- \int\ d{\bf y}\
e^{-i(\sigma x_\pom P_{\!A}^+y^- - {\bf s}\cdot {\bf y})}
\nonumber\\
&& \times
\langle P_{\!A^\prime},s_{\!A^\prime}|A_a^{\nu}(0,y^-,{\bf y})\,
A_a^{\mu}(0)|P_{\!A},s_{\!A} \rangle.
\label{Smunu}
\end{eqnarray}
The function $S$ is simply related to the operator matrix element
$A^{ij}$, Eq.~(\ref{Aij}), that appeared in our discussion of the
diffractive gluon distribution:
\begin{equation}
\int { d{\bf s} \over (2\pi)^2 }\
S^{ij}(\sigma,{\bf s};x_\pom,t;s_{\!A},s_{\!A^\prime})
= A^{ij}(\sigma,x_\pom,t;s_{\!A},s_{\!A^\prime}).
\end{equation}
The function
\begin{equation}
{\cal M}_{IJ}^{\mu\nu}({\bf q},{\bf k},{\bf s},\sigma,\tau;s_q,s_k)
\end{equation}
represents the amputated hard interaction graph.  Here ${\bf q}$ is
the momentum of the antiquark that enters the final state and ${\bf
k}$ is the transverse momentum of the probed quark. We have ${\bf k}
= -{\bf q}-{\bf l}$ by momentum conservation.  The variables $s_k$
and $s_q$ are the helicities of the probed and final state quarks
respectively; $I$ and $J$ are their colors.

There is an important difference with the gluon case. The quark has a
mass $m_q$. Thus the minus momentum of the on-shell quark entering
the final state is
\begin{equation}
q^- = {({\bf q^2} + m_q^2)\over 2 \tau x_\pom P_{\!A}^+}.
\end{equation}
Then
\begin{equation}
k^2 \approx {({\bf q^2} + m_q^2)\over 2 \tau }
\label{ksqquark}
\end{equation}
in the $\tau \to 0$ limit. What counts here is the mass of the final
state quark as it emerges from the hard interaction and propagates
into the final state. Presumably the best model for $m_q$ in this
role is the constituent quark mass ($\sim 0.3\ {\rm GeV}$), not the
much smaller current quark mass. This is a substantial mass, so that
the condition that defined whether the virtual lines in $\cal M$ are
far off shell,
\begin{equation}
{({\bf q^2} + m_q^2)\over \tau } \gg m^2,
\end{equation}
is satisfied for any ${\bf q^2}$ when $\tau$ is small. Thus we do not
need to restrict the integration region. On the other hand, the region
${\bf q^2} \ll m_q^2$ is not important in the integration.

We evaluate $\cal M$ in the $\tau \to 0$ limit using null plane spin
defined with the plus direction as special for the probed quark and
defined  with the minus direction as special for the final-state
antiquark. We find
\begin{equation}
{\cal M}^{\mu\nu}
\approx
{C_F \over 8}\, g^2\, \delta_{IJ}\,
{\sqrt \tau \over \sqrt {{\bf q}^2  + m_q^2}}\
w(s_k)^\dagger\, \Gamma^{\mu\nu}\, w(-s_q).
\label{calMquark}
\end{equation}
The $w(s)$ are two component spinors $w(+{\scriptstyle{1\over 2}}) =
(1,0)$ and  $w(-{\scriptstyle{1\over 2}}) = (0,1)$. Then we can write
the $2\times 2$ matrix $\Gamma$ using transverse Pauli spin matrices
$\sigma^1$ and $\sigma^2$.  For transverse indices $\mu\nu$, we find
\begin{eqnarray}
\Gamma^{ij}&=&
{1 \over \sigma + i\epsilon}\, \sigma^i ({\bf s}\cdot \sigma) \sigma^j
+{1 \over 1+\sigma}\, \sigma^j ({\bf r}\cdot \sigma) \sigma^i
\nonumber\\
&&
+ 2\delta^{ij} ({\bf k}\cdot \sigma)
+ {2 \over \sigma + i\epsilon}\, \sigma^i q^j
- {2 \over 1+\sigma}\, \sigma^j q^i
\nonumber\\
&&
+i\left(
-{1 \over \sigma + i\epsilon}\sigma^i\sigma^j
+{1 \over 1+\sigma}\sigma^j\sigma^i
+2 \delta^{ij}
\right)\
m_q
\label{Gammaij}
\end{eqnarray}
(We hope that the Pauli spin matrices $\sigma^i$ will not be confused
with the momentum fraction $\sigma$ that occurs in this equation in the
combinations $1/\sigma$ and $1/(1+\sigma)$.) For one transverse index
and one plus index we find
\begin{equation}
{1 \over k^+}\,\Gamma^{+j} \approx 2 \sigma^j
\label{Gamma+j}
\end{equation}
Also ${\Gamma}^{i+} \approx {\Gamma}^{+i}$, while  ${\Gamma}^{++}$ does
not give leading contributions as $\tau \to 0$. Finally, we note that
${\Gamma}^{-\nu}$ and ${\Gamma}^{\mu -}$ are not needed because they
multiply 0 in $A^+ = 0$ gauge.

The spin function $\Gamma^{\mu\nu}$ is rather complicated, but we are
concerned with only two of its properties. First, if we think of
$\Gamma$ in coordinate space as a function of the separation $y^\mu$
between the points where the two exchanged gluons attach, then $\Gamma$
is proportional to $\delta(y^+)$ and to a linear combination of
$\delta(y_T)$ and $\partial\delta(y_T)/\partial y_k$. Thus the
interaction is hard in the sense of being local in $y^+$ and $y_T$.
Second, and most important, $\Gamma$ is independent of $\tau$.

We can now insert Eqs.~(\ref{ksqquark}) and (\ref{calMquark}) into
Eqs.~(\ref{diffintquark}) to obtain the $\tau$ dependence of the
diffractive quark distribution:
\begin{eqnarray}
df_{q/A}^{\rm diff} \over {dx_\pom\,dt} &=&
\tau^2\
{1 \over {64\pi^3}}\,
{ 3 C_F^2 g^4 \over 16}\,
\,{ 1 \over 2}\sum_{s_{\!A},s_{\!A^\prime}}\,
\int d\sigma d\sigma^\prime
\int {{d^2{\bf s}} \over {(2\pi)^2}}\
\int {{d^2{\bf s}^\prime} \over {(2\pi)^2}}\
\nonumber\\
&&\times
S_{\rho\sigma}^*(\sigma^\prime,{\bf s}^\prime;x_\pom,t;
s_{\!A},s_{\!A^\prime})
S_{\mu\nu}(\sigma,{\bf s};x_\pom,t;s_{\!A},s_{\!A^\prime})
\nonumber\\
&&\times
\int {{d^2{\bf q}} \over {(2\pi)^2}}\
{{\rm Tr}\left\{
\Gamma^{\rho\sigma}(\sigma^\prime,{\bf s}^\prime,{\bf q})^\dagger
\Gamma^{\mu\nu}(\sigma,{\bf s},{\bf q})
\right\}
\over ({\bf q}^2 + m_q^2)^3 }.
\label{diffquarkresult}
\end{eqnarray}
The crucial feature here is the factor of $\tau^2$.

We conclude that the constituent counting result for the diffractive
distribution of quarks is
\begin{equation}
{df_{q/A}^{\rm diff}(\beta x_\pom,\mu;x_\pom,t) \over {dx_\pom\,dt}}
\propto (1-\beta)^2.
\end{equation}
However, suppose that we interpret the calculation of the previous
section as saying that the diffractive distribution of gluons is
proportional to $(1-\beta)^0$ for $\beta$ near 1 when the scale $\mu$ is not
too large. Then the evolution equation for the diffractive parton
distributions will give a quark distribution that behaves like
\begin{equation}
{df_{q/A}^{\rm diff}(\beta x_\pom,\mu;x_\pom,t) \over {dx_\pom\,dt}}
\propto (1-\beta)^1.
\end{equation}
when the scale $\mu$ is large enough that some gluon to quark
evolution has occurred, but not so large that effective power $p$ in
$(1-\beta)^p$ for the gluon distribution has evolved substantially
from $p = 0$. A signature of this phenomenon is that the diffractive
quark distribution will be growing as $\mu$ increases at large
$\beta$, rather than shrinking. Perhaps this is seen in the data
\cite{ZeusH1}.

\section{Conclusion}
\label{Conclusion}

We close with some observations concerning the implications for
HERA physics of the discussion presented here.

We have discussed two kinds of factorization relevant to diffractive
hard scattering. Both are experimentally testable. We say that
{\it diffractive factorization} holds if the cross section is
a standard partonic hard scattering cross section convoluted with a
diffractive parton distribution.  The diffractive parton distribution
gives the distribution of partons in the hadron under the condition
that the hadron is diffractively scattered. (If there is a second
hadron in the initial state and we do not demand that it be
diffractively scattered, then the cross section should also contain a
convolution with the  ordinary parton distribution within that
hadron.) We say that {\it Regge factorization} holds if the
diffractive parton distribution is a product of
$x_\pom^{-2\alpha(t)}$ times a pomeron-hadron coupling times a
function $f_{a/\pom}(\beta,t,\mu)$ that is interpreted as the
distribution of partons ``in'' the pomeron. These properties together
constitute the Ingelman-Schlein model \cite{IS}.

Diffractive factorization is something that can be disproved for a
given process by a counterexample at some fixed order of perturbation
theory (with wave functions for the hadronic states).
Correspondingly, it could in principle be proved to hold at any fixed
order of perturbation theory, which one would take as a strong
indication that it holds beyond perturbation theory for that process.
Diffractive factorization makes no statement about the Regge
structure of the nonperturbative factors. Regge factorization does
make such a statement. It would be interesting to study Regge
factorization from the point of view of the BFKL pomeron, perhaps
extracting a model for the distribution of partons in the pomeron.

Let us discuss first diffractive deeply inelastic scattering, which is
simpler than diffractive hard scattering processes with two hadrons
in the initial state. For diffractive deeply inelastic
scattering, we argue that diffractive factorization is a
consequence of perturbative QCD, although a detailed proof is beyond
the scope of this paper. From the point of view of current theory,
Regge factorization for the diffractive parton distributions is a
conjecture based on experience with soft diffractive scattering.

The HERA experiments \cite{ZeusH1} have now provided evidence
concerning these issues. It is a consequence of diffractive
factorization that the diffractive structure function $F_2^{\rm
diff}$ should exhibit approximate scaling as $Q^2$ is increased with
fixed $x_\pom$ and $\beta$. This is confirmed by the data. The
dependence on $x_\pom$ in the form $x_\pom^{1-2\alpha(t)}$ predicted
by Regge factorization, Eq.~(\ref{IS}), is also confirmed by the data.

For the future, it will be important to measure the diffractive
parton distributions in as complete detail as possible, using charged
and neutral current events, $F_2^{\rm diff}$ and $F_3^{\rm diff}$,
and probes for heavy flavors in the final state \cite{CTEQpom}.
Especially crucial is the diffractive gluon distribution. This can be
measured using diffractive deeply inelastic scattering with high
$P_T$ jets detected in the final state. If we demand that we see two
jets instead of the usual single struck quark jet, and if these two
jets have a high transverse momentum relative to the direction
determined by the sum of their momenta, then the hard process is of
order $\alpha\alpha_s$ instead of just $\alpha$. For such a process,
gluons participate as initial partons on the same footing as quarks.
Since the diffractive quark distributions are already known, it
should be possible to extract the diffractive gluon distributions.

Our analysis suggests that the diffractive gluon distribution is
quite hard, with a behavior between $(1-\beta)^1$ and $(1-\beta)^0$
for large $\beta = \xi/x_\pom$ at moderate values of the scaling
parameter $\mu$, say 2 GeV. The corresponding behavior of the
diffractive quark distribution, which is quite directly measured in
$F_2^{\rm diff}$, is between $(1-\beta)^2$ and $(1-\beta)^1$. Here the
$(1-\beta)^1$ for quarks would arise if the diffractive gluon
distribution is large and behaves like $(1-\beta)^0$, so that the
quarks at large $\beta$ are produced by $g \to q + \bar q$.

Given a complete set of diffractive parton distributions, it will be
interesting to test the evolution equation (\ref{APeqn}).

Let us now turn to hard processes with two hadrons in the initial
state, as exemplified by $\gamma + p \to jets + p + X$ at HERA, where
we look at the hadronic part of the photon. In this case, the
factorization that holds in {\it inclusive} hard scattering is
expected to break down in {\it diffractive} hard scattering, as
shown by counterexamples at a fixed order of perturbation
theory \cite{CFS,BS}. If one extracts diffractive parton distribution
functions from deeply inelastic scattering and uses them to predict
cross sections for $\gamma + p \to jets + p + X$, then the observed
cross section should contain extra terms that do not match the
prediction \cite{BS}. In particular, there should be extra
contributions that correspond to the jets carrying almost all of the
longitudinal momentum of the pomeron. Perhaps this corresponds to the
``superhard'' component seen in the UA8 experiment \cite{UA8} in $\bar
p + p \to p + jets + X$.

In summary, the HERA results \cite{ZeusH1}, together with the
earlier UA8 results \cite{UA8}, have confirmed the basic features of
the Ingelman-Schlein picture of diffractive deeply inelastic
scattering. The experiments have shown that diffractive scattering is
related to exchanges of quanta that, when examined with a hard probe,
appear to be the pointlike quarks and gluons of QCD. Much more
remains to be done, but already we are challenged to connect the
theory of pointlike gluons to the soft color dynamics that is
presumably responsible for diffractive scattering.

\acknowledgments

We thank G.\ J.\ van Oldenborgh for explicitly rechecking his
algorithms in some of the sensitive low $t$ regions investigated in
Appendix \ref{Appendix:Model}. We thank J.\ C.\ Collins, G.\ Ingelman,
J.\ Bartels and many members of the H1 and Zeus collaborations for
helpful conversations.

\appendix
\section{Structure of the hard subgraph}
\label{Appendix:HardStructure}

In this appendix we investigate the structure of the hard subgraph
for the diffractive distribution of gluons in hadron in the limit
$\xi/x_\pom \to 1$. Recall that the hard subgraph is a function of
momenta $r^\mu, s^\mu$, and $q^\mu$, where $r^\mu$ and $s^\mu$ are
the momenta of the gluons exchanged with the soft subgraph and
$q^\mu$ is the momentum of the gluon that goes into the final
state. Since $q^\mu q_\mu = 0$ we will consider the hard
subgraph to be a function of ${\bf q}$ and $q^-$, replacing $q^+$ by
${\bf q}^2/(2 q^-)$. The momentum of the detected gluon is $k^\mu = r^\mu -
s^\mu - q^\mu$.

In Sec.~\ref{TauQuark}, we studied the hard subgraph in the limit
that applies when $\xi/x_\pom \to 1$. This limit is really a dual
limit. We take $r^+ \sim s^+ \sim x_\pom P_{\!A}^+$ but
\begin{equation}
{ q^+ \over s^+}\equiv {{\bf q}^2 \over 2 s^+ q^-}\ll 1.
\label{limitqplus}
\end{equation}
We also recall the definition (\ref{calR}) of the integration region
considered for ${\bf q}$,
\begin{equation}
{ {\bf q}^2 \over \tau} \gg m^2.
\end{equation}
That is, $q^-\gg m^2/(2x_\pom P_{\!A}^+)$. We combine this with the
assumption that, in the effective integration region for ${\bf r}$,
${\bf s}$, $r^-$ and $s^-$, these variables have values typical for
slow gluons:
\begin{equation}
r^- \sim s^- \sim {m^2 \over 2x_\pom P_{\!A}^+}.\quad
{\bf r}^2 \sim {\bf s}^2 \sim m^2.
\end{equation}
Then
\begin{equation}
{ r^- \over q^-}\ll 1,\quad\quad
{ s^- \over q^-}\ll 1,\quad\quad
{ {\bf r}^2 \over 2 s^+ q^-}\ll 1,\quad\quad
{ {\bf s}^2 \over 2 s^+ q^-}\ll 1.
\label{limitqminus}
\end{equation}

We claimed in Sec.~\ref{decompose} that in this limit, the hard
subgraph is independent of the variables ${\bf r}$, ${\bf s}$,
$r^-$, and $s^-$. We also claimed that the transverse components of
the hard subgraph dominate over other components in the limit
considered, as in Eq.~(\ref{HardSoftindices}). In this appendix, we
substantiate these claims.

We consider first the question of the independence on the variables
${\bf r}$, ${\bf s}$, $r^-$, and $s^-$, taking, for the moment, only
the transverse components of the hard scattering subgraph. This is
the same as multiplying the hard subgraph by purely transverse
polarization vectors for the two gluons exchanged with the soft
subgraph.

We consider the hard amplitude $\cal M$, defined as in
Sec.~\ref{hard}. Thus $\cal M$ does not include the propagator for
the detected gluon with momentum $k^\mu$, but does include an
$\epsilon^+ = 0$ gauge polarization vector for this gluon. It also
includes an $\epsilon^- = 0$ gauge polarization vector for the gluon
with momentum $q^\mu$ that enters the final state.

As a matter of convenience, we will analyze the first graph in
Fig.~\ref{hardfig}. Essentially the same argument covers the second
graph, while the third graph, with a four gluon interaction, is quite
trivial.

The Feynman rules give $\cal M$ as a rational function of the
components of  $r^\mu, s^\mu$, and $q^\mu$. We are interested in
particular in the behavior of $\cal M$ for small values of $r^-$,
$s^-$, ${\bf r}$, ${\bf s}$ and ${\bf q}$ as specified in
Eqs.~(\ref{limitqplus}) and (\ref{limitqminus}). For this reason, we
need to know if there are any factors of the small variables in the
denominator of $\cal M$.

The gluon propagator in Fig.~\ref{hardfig} is
\begin{equation}
{i D^{\mu \nu}(q+s) \over (q+s)^2 }=
i{-g^{\mu\nu} (s^+ + q^+)
+ { (s+q)^\mu \delta_-^\nu +  \delta_-^\mu (s+q)^\nu}
\over
(s^+ + q^+)\,
[2 ( s^+ + q^+) (q^- + s^-)
- ({\bf s} + {\bf q})^2] }
\end{equation}
Setting $q^+ = {\bf q}^2/(2q^-)$, the denominator becomes
\begin{eqnarray}
&&2 (s^+)^2 q^-\
\left(1 + {{\bf q}^2 \over 2 s^+ q^-}\right)\
\biggl\{
\left(1 + {{\bf q}^2 \over 2 s^+ q^-}\right)\,
\left(1 + {s^- \over q^-}\right)
-{{(\bf s} + {\bf q})^2 \over 2 s^+ q^-}
\biggr\}.
\nonumber\\
\end{eqnarray}
Then using Eqs.~(\ref{limitqplus}) and (\ref{limitqminus}) (in either
order), the denominator becomes
\begin{equation}
 2 (s^+)^2 q^-
\end{equation}
That is, the denominator does not contain factors of the small
variables, so that it has a finite limit as the small variables tend
to zero.

We can write $\cal M$ as a product
\begin{equation}
{\cal M}^{ijmn} =
\epsilon(s,i)_\alpha\,
\epsilon(r,j)_\beta\,
\epsilon(k,m)_\gamma\,
\epsilon(q,n)_\delta\,
M^{\alpha\beta\gamma\delta}.
\end{equation}
The transverse components of the polarization vectors have the form
\begin{equation}
\epsilon(p,i)^I = \delta_{iI}
\end{equation}
for $p^\mu$ stands for any of the momenta $r^\mu$, $s^\mu$, $q^\mu$ or
$k^\mu$. For $p^\mu = r^\mu$ or $s^\mu$ we are (temporarily) defining
$\epsilon(p,i)^+ = \epsilon(p,i)^- = 0$. The polarization vector for
the detected gluon has $\epsilon(k,m)^+ = 0$ but has a non-zero minus
component
\begin{equation}
\epsilon(k,m)^- = { k^m \over k^+}
= {r^m - s^m - q^m \over r^+ - s^+ - q^+}.
\label{polarizationk}
\end{equation}
The polarization vector for the final-state gluon, which we take to
be in $\epsilon^- = 0$ gauge according to Eq.~(\ref{Aminusgauge}),
has a non-zero plus component
\begin{equation}
\epsilon(q,n)^+ = { q^n \over q^-}.
\label{polarizationq}
\end{equation}
Thus neither the gluon propagator nor any of the polarization
vectors contains a factor of a small variable in the denominator.

We now consider the transverse tensor
$
{\cal M}^{ijmn}
$
as a function of the variables
\begin{equation}
\{r^+,s^+,r^-,s^-,q^-,{\bf r},{\bf s},{\bf q}\}.
\end{equation}
${\cal M}$ is a rational function of these arguments and has dimension
$D = 0$ and boost dimension $B = 0$, where $B$ gives the scaling
under boosts in the $z$ direction. That is, for any four-vector
$p^\mu$,
\begin{eqnarray}
p^+ &\quad{\rm has}\quad&  D=1,\ B = 1,
\nonumber\\
|{\bf p}| &\quad{\rm has}\quad&  D=1,\  B = 0,
\nonumber\\
p^- &\quad{\rm has}\quad&  D=1,\  B = -1.
\end{eqnarray}
Since ${\cal M}$ has $B=0$ and $D = 0$, it can be written as a
function of a reduced number of arguments, each of which has $B =
D = 0$:
\begin{equation}
{\cal M}^{ijmn}\!\left(
{ r^+ \over s^+},
{r^- \over q^-},
{s^- \over q^-},
{{\bf r} \over \sqrt{s^+q^-}},
{{\bf s} \over \sqrt{s^+q^-}},
{{\bf q} \over \sqrt{s^+q^-}}
\right).
\end{equation}
We are interested in the limit in which the last five arguments are
small, and we know that none of these arguments occur as factors in
the denominators. Thus when  Eqs.~(\ref{limitqplus}) and
(\ref{limitqminus}) hold, ${\cal M}$ approaches the limiting
value
\begin{equation}
{\cal M}^{ijmn}\!\left(
{ r^+ \over s^+},0,0,{\bf 0},{\bf 0},{\bf 0}
\right).
\end{equation}

We thus confirm the claim made in Sec.~\ref{decompose} that we can
neglect ${\bf r}$, ${\bf s}$, $r^-$, and $s^-$ in $\cal M$.
Furthermore, since the limiting function is covariant under rotations
about the $z$ axis, it must be a linear combination of the tensors
$\delta^{ij}\delta^{mn}$, $\delta^{im}\delta^{jn}$ and
$\delta^{in}\delta^{jm}$. Each of these tensors multiplies a rational
function of $r^+/s^+ = (1+\sigma)/\sigma$. That is, the coefficient
functions are rational functions of $\sigma$. This is, of course, just
the structure given in Eq.~(\ref{calMresult}). The argument given
above does not establish that the limiting function is nonzero, but
this is what we found by calculation.

We now return to the issue of whether the transverse components of
the hard subgraph dominate over other components, as claimed in
Sec.~\ref{decompose}. We consider
\begin{equation}
\tilde{\cal M}^{\alpha\beta mn} =
\epsilon(k,m)_\gamma\,
\epsilon(q,n)_\delta\,
M^{\alpha\beta\gamma\delta}
\end{equation}
and choose values other than $\left\{1,2\right\}$ for $\alpha$
or $\beta$ or both. Now $\alpha = -$ or $\beta = -$ are not
possible: $\tilde{\cal M}^{\alpha\beta mn}$ multiplies the
soft subgraph with corresponding indices, call it ${\cal
S}_{\alpha\beta}$, which vanishes for $\alpha = -$ or $\beta = -$
because of the gauge condition $A^+(x) = 0$ (that is, $A_-(x) = 0$).
Thus we should consider $\alpha = +$ or $\beta = +$. Let us
consider $\alpha = +$ with $\beta = j \in\{1,2\}$ as an example that
illustrates the general argument. Thus we wish to investigate
whether $\tilde{\cal M}^{+jmn} {\cal S}_{+j}$ is dominated by
$\tilde{\cal M}^{ijmn} {\cal S}_{ij}$ in the limit specified by
Eqs.~(\ref{limitqplus}) and (\ref{limitqminus}). We need an order of
magnitude estimate for the ratio of ${\cal S}_{+j}$ to
${\cal S}_{ij}$. An analysis of lowest order graphs indicates that
this is the same as the ratio of the corresponding components of
the polarization vectors for a typical slow gluon,
$\epsilon^-(s^\mu,i)/ \epsilon^j(s^\mu,i) \sim m/(x_\pom P_{\!A}^+)$. We
write this as
\begin{equation}
{ {\cal S}_{+j} \over {\cal S}_{ij}}
\sim { |{\bf s}| \over s^+}
\end{equation}
and compare
\begin{equation}
{ |{\bf s}| \over s^+}\tilde{\cal M}^{+jmn}
\end{equation}
to $\tilde{\cal M}^{ijmn}$. The analysis is simple. $\tilde{\cal
M}^{+jmn}$ has dimension $D = 0$ and boost dimension $B = 1$. Thus
it can be written as $\sqrt{s^+/q^-}$ times a function
${\cal N}^{jmn}$ that has dimension $D = 0$ and boost dimension $B =
0$ and, like ${\cal M}^{ijmn}$, has no factors of the small
variables in its denominator. As our previous analysis shows,
${\cal N}^{jmn}$ has a finite limit as the the small variables tend
to zero. (Actually, ${\cal N}^{jmn}$ vanishes in this limit because
of its spin structure, but we will not need to use this fact.)
We note that the factor that multiplies ${\cal N}^{jmn}$
\begin{equation}
{ |{\bf s}| \over s^+}\,\sqrt{s^+\over q^-}
= { |{\bf s}| \over \sqrt{s^+ q^-}}
\end{equation}
vanishes in the limit specified by Eq.~(\ref{limitqminus}). This
establishes the claim.

\section{A model}
\label{Appendix:Model}

In this appendix we compute the diffractive gluon distribution
${{df^{\rm diff}_{g/A}}/ {dx_\pom\,dt}}$ in a simple model. As
we will see, this model does not exhibit the pomeron behavior
$x_\pom^{- 2\alpha(t)}$ with $\alpha(t) \approx 1$. Nevertheless, the
model provides a check that
$
{{df^{\rm diff}_{g/A}(\beta x_\pom;x_\pom,t)}/{dx_\pom\,dt}}
$
can have a finite limit as $\beta  \to 1$ in an exact numerical
evaluation of the graphs to a given order of perturbation theory.

The model we will use is scalar-quark QCD, given by the Lagrangian,
\begin{eqnarray}
{\cal L} &=&
D_{\mu}\bar{q}\,D^{\mu}q
- m^2\bar{q}q
- {1\over 4} G^{a}_{\mu \nu } G^{\mu \nu}_a
- {1\over 4}g_4\,(\bar{q}q)^2
-{1\over 2\xi}
(\partial \cdot A_a)^2
+ \left\{\hbox{\rm Faddeev-Popov}\atop{\rm terms}
\right\}
\nonumber\\
&& + {1\over 2}(\partial \phi) ^2
- {1\over 2}M^2\phi ^2
- G\phi \bar{q}q
\nonumber\\
&=&
{\cal L}_{\rm QCD}+{\cal L}_{\phi}.
\label{lagran}
\end{eqnarray}
This model has soft and collinear singularities just as in QCD. As
one further simplification, we model the diffractively scattered
particle by the scalar $ \phi $ field with a $\phi \bar q q$
interaction to the quarks. Then the perturbative $\phi
\bar q q$ interaction plays the role of the  nonperturbative
Bethe-Salpeter wave function of a real QCD meson. The probability for
finding a $q\bar q$-pair inside the meson in this model falls of as
$1/{\bf k}^4$ in the ultraviolet, where $\bf k$ is the transverse
momentum of the quark. This good behavior in the ultraviolet is due
to the fact that G carries the dimension of mass.  Thus the model
allows for a simple treatment that approximately simulates
properties of bound quarks inside a hadron.

\begin{figure}[htb]
\centerline{ \DESepsf(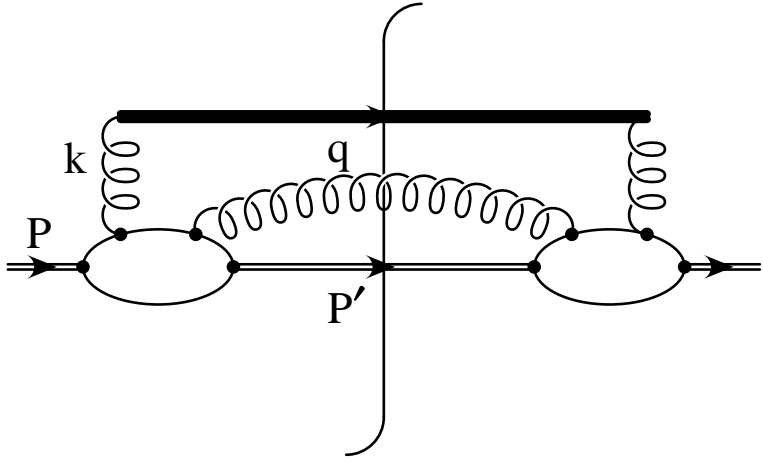 width 10 cm) }
\vskip 0.5 cm
\caption{
Diagram contributing to the diffractive gluon distribution in a
meson in the model of this appendix.}
\label{figmod}
\end{figure}

\begin{figure}[htb]
\centerline{ \DESepsf(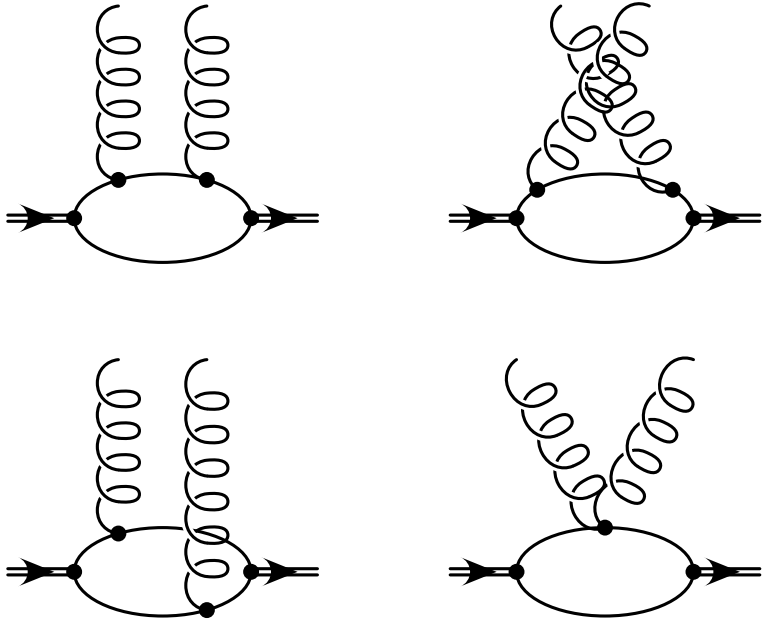 width 10 cm) }
\vskip 0.5 cm
\caption{
Four  graphs for the amplitude in Fig.~(\protect\ref{figmod})}
\label{figblobs}
\end{figure}

Using this model, we compute the diffractive gluon distribution
function within a meson, $ d f^{{\rm diff}}_{g/A}(\beta
x_\pom;x_\pom,t) / {dx_\pom\,dt}$, as given in Eq.~(\ref {fdifftoG}),
to the lowest nontrivial order in $g$, order $g^4$. One of the order
$g^4$ diagrams for the function $G_{g/A}^{{\rm diff}}(P,q,k)$ in
Eq.~(\ref {fdifftoG}) is shown in Fig.~\ref{figmod}. One gluon is
detected, and  a single gluon is exchanged between the two quark
loops on opposite sides of the final state cut.  This gluon carries
that portion of the momentum transfer from the meson that is lost to
the detected gluon (and in a physical process would be lost to the
hard interaction).  The heavy bar ending with gluon lines represents
the gluon density operator in Eq.~(\ref{fdiff1}).  In addition to the
graph shown, at each loop the gluons can also attach to the lower
quark line.  Also in both cases there is a graph where the gluon
lines are crossed.  In addition one gluon can attach on each of the
quark lines.  Finally there are the two-gluon two-quark contact
interaction graphs. In total this implies $8^2 = 64$ combinations.
After symmetry considerations, there are four types of amplitudes
that must be explicitly evaluated.  The four amplitudes are shown in
Fig.~\ref{figblobs}.

The final form for $G_{g/A}^{{\rm diff}}$ that we obtain is,
\begin{eqnarray}
G_{g/A}^{{\rm diff}}(P,q,k) &=&
{1\over {2(2\pi)^3\beta(1-\beta)x_{\pom}^2}}\int d^2 {\bf q}\
{{G^{\mu\alpha}(P,q,k)^*\
G^{\nu\beta}(P,q,k)\
g_{\alpha\beta}\,c_{\mu\nu}(k)}
\over {(k^2+i\epsilon)(k^2-i\epsilon)}}
\label{pertdist}
\end{eqnarray}
where $k^+=\beta x_{\pom}P^+$,
\begin{eqnarray}
c_{\mu\nu}(k) &=&
\frac{1}{(P^+)^2}\
(k^{+2} g_{\mu\nu}
+ k^2 g_{\mu}^+g_{\nu}^+
- k_{\mu} k^+ g_{\nu}^+
- k_{\nu} k^+ g_{\mu}^+),
\end{eqnarray}
and $G^{\mu\nu}(P,q,k)$ is a second rank tensor containing the sum of
quark loops with all possible combinations of attachments by two
gluons.

We have evaluated the loop integrals in Eq.~(\ref{pertdist}) in
terms of their explicit Spence function expressions.  For this we
have used the computer code of G.\ J.\ van Oldenborgh and J.\ A.\ M.\
Vermaseren \cite{gj}. Their method is an independently formulated
extension of the Form algorithm of G.\ 't Hooft and M.\ Veltman
\cite{thooft}.  The authors of \cite{gj} claim that their algorithms
provide easier isolation of both asymptotic behavior and potential
numerical instabilities. In one of the most sensitive regions, that
of low $t$, their algorithms have proven to be more accurate.

To summarize the specifications of our calculation, it was for
small $t$, one external line was massless, and one had to integrate
over a large region in the transverse space of the variable ${\bf q}$.
Due to gauge invariance, several constraints are placed on
$G^{\mu\nu}$ which we have numerically checked to hold.
$G^{\mu\nu}(P,q,k)$ has the form
\begin{equation}
G^{\mu\nu}(P,q,k) = \sum_{J = 1}^5 T_J^{\mu \nu} A_J\,.
\end{equation}
Here the $T_J$ are tensors made from the available vectors $P,q,k$.
The $A_J$ are scalar functions of the dot products of these momenta.
In general there would be ten terms in the sum, but gauge invariance
cuts this in half. To find the possible $T_J$, define
\begin{eqnarray}
P_q^\mu &=& q^2\, P^\mu - P\cdot q\, q^\mu\,,
\nonumber\\
P_k^\nu &=& k^2\, P^\nu - P\cdot k\, k^\nu\,,
\nonumber\\
S_q^\mu &=& k\cdot q\, P^\mu - P\cdot q\, k^\mu\,,
\nonumber\\
S_k^\nu &=& q\cdot k\, P^\nu - P\cdot k\, q^\nu\,.
\end{eqnarray}
Notice that $q\cdot P_q = 0$, {\it etc}.  Now define
\begin{eqnarray}
T_1^{\mu\nu} &=& P_q^\mu P_k^\nu\,,
\nonumber\\
T_2^{\mu\nu} &=& P_q^\mu S_k^\nu\,,
\nonumber\\
T_3^{\mu\nu} &=& S_q^\mu P_k^\nu\,,
\nonumber\\
T_4^{\mu\nu} &=& S_q^\mu S_k^\nu\,,
\nonumber\\
T_5^{\mu\nu} &=& q\cdot k\, g^{\mu\nu} - k^\mu q^\nu\,.
\label {gauge1}
\end{eqnarray}
These have the property that they are polynomials in the components
of the momenta and that $q_\mu T_J^{\mu\nu} = 0$ and
$k_\nu T_J^{\mu\nu} = 0$.  Thus $G^{\mu\nu}$ is properly gauge
invariant. In our calculation, we transformed $G^{\mu\nu}$ into a
basis where the first five tensor forms are those given above and the
other five are
\begin{eqnarray}
T_6^{\mu\nu} &=& P_k^\mu P_q^\nu\,,
\nonumber\\
T_7^{\mu\nu} &=& P_k^\mu S_q^\nu\,,
\nonumber\\
T_8^{\mu\nu} &=& S_k^\mu P_q^\nu\,,
\nonumber\\
T_9^{\mu\nu} &=& S_k^\mu S_q^\nu\,,
\nonumber\\
T_{10}^{\mu\nu} &=&  g^{\mu\nu} .
\label {gauge2}
\end{eqnarray}
We have checked over a wide range of the parameters $x_\pom$, $t$
and the transverse momentum ${\bf q}$ that the coefficients
of the tensors in Eq.~(\ref {gauge2}) vanish, leaving only
the tensors in Eq.~(\ref {gauge1}) with nonvanishing coefficients.
As a second check for gauge invariance of $G^{\mu\nu}$, we have
numerically checked that $q_\mu G^{\mu\nu} = 0$ and $k_\nu G^{\mu\nu}
= 0$.

\begin{figure}[htb]
\centerline{ \DESepsf(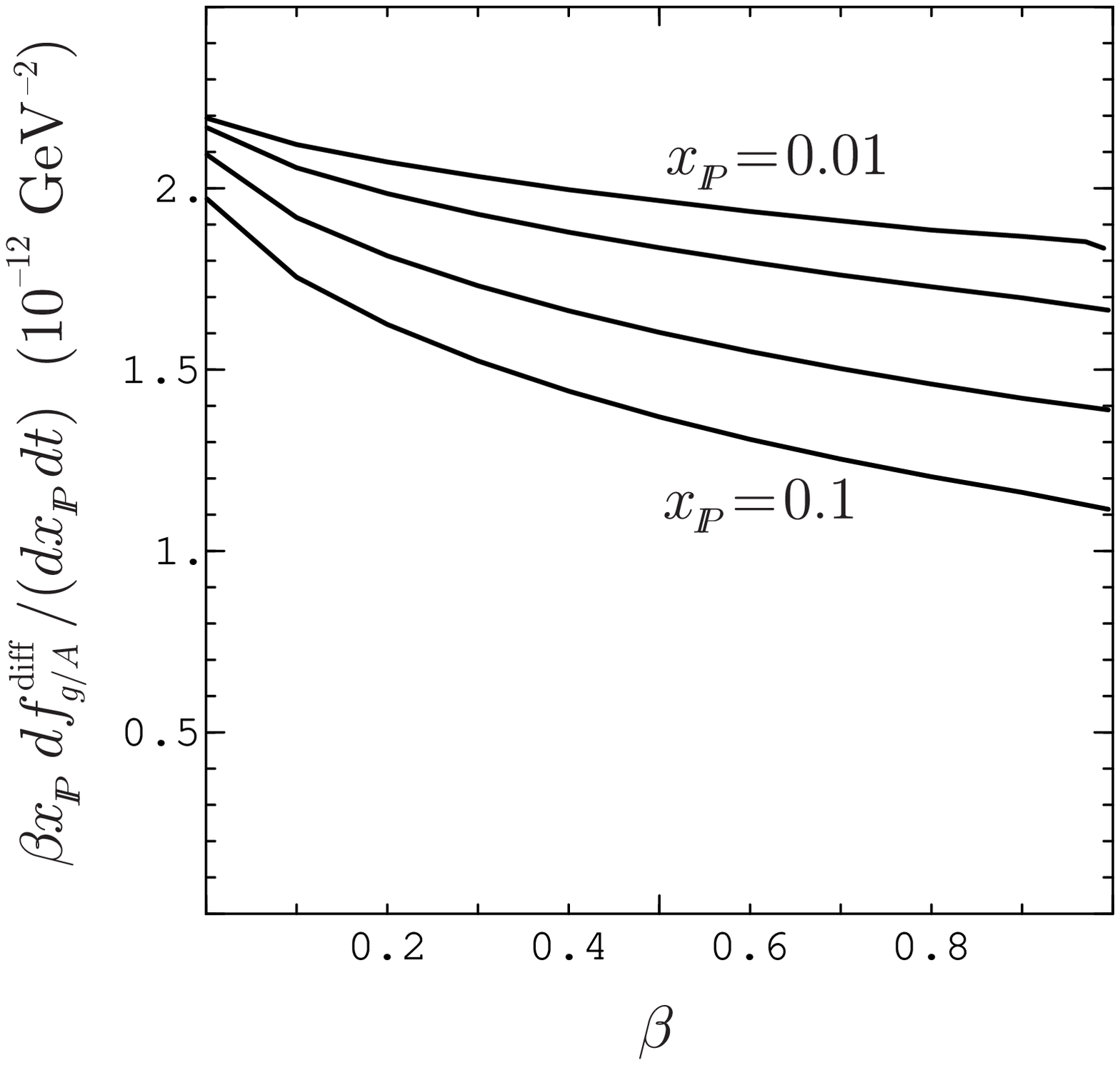 width 12 cm) }
\vskip 0.5 cm
\caption{
Diffractive gluon distribution $\beta x_\pom\, df_{g/A}^{\rm
diff}(\beta x_\pom;x_\pom,t)/ dx_\pom\,dt $ in the model of this
appendix as a function of $\beta$ for (from top to bottom)
$x_\pom$=0.01, 0.02, 0.05, 0.1 with $|t| = 1\ {\rm GeV}$ In this
model, there is no dependence on a renormalization scale $\mu$. The
parameter choices are $M^2 = m^2=0.1{\ \rm GeV}^2$, $G= 0.3 {\ \rm
GeV}$ and $g = 0.1$ }
\label{figcurves}
\end{figure}

In Fig.~\ref{figcurves} we show the diffractive gluon distribution
function multiplied by $\beta x_\pom$, at $|t|=1 {\ \rm GeV}^2$, and
for a selected choice of $x_\pom$ values. The masses of the quarks and
mesons were  $M^2 = m^2=0.1{\ \rm GeV}^2$, while the couplings were
$G = 0.3 {\ \rm GeV}$ and $g = 0.1$.  We see that the model does {\it
not} exhibit the behavior
\begin{equation}
{{df^{\rm diff}_{g/A}(\beta x_\pom; x_\pom,t)}\over{dx_\pom\,dt}}
\propto x_\pom^{-2}
\end{equation}
for $x_\pom \to 1$ at fixed $\beta$ that would be characteristic of
pomeron exchange. One needs at least one rung of a gluon ladder to
obtain this behavior.  The model exhibits the behavior
\begin{equation}
{{df^{\rm diff}_{g/A}(\beta x_\pom; x_\pom,t)}\over{dx_\pom\,dt}}
\propto (1-\beta)^0
\end{equation}
as $\beta \to 1$ at fixed $x_\pom$, as in Eq.~(\ref{ginpomresult}).

One can understand the $\beta$ and $x_{\pom}$ behavior seen in this
numerical study from an analytic viewpoint.  In Sec.~\ref{TauGlue},
the detected and emitted gluons coupled do a slow gluon. In the
present simple model, they  couple directly to a fast quark. This
means that the virtual quark line in Fig.~\ref{figmod} is far off
shell when ${\bf q}^2/((1-\beta) x_\pom) \gg m^2$. For example, it is
off shell for ${\bf q}^2 \sim m^2$ for any $\beta$ as long as we take
$x_\pom \ll 1$. This gives quite a different behavior from that found
in Sec.~\ref{TauGlue}. In the model, the loop serves as a built-in
low momentum cut-off. However, the meson can be replaced by a single
fast quark line if we substitute an artificial low momentum cut-off
\begin{equation}
{\bf q}^2 >  (1-\beta) x_\pom \ m^2
\label{modq2}
\end{equation}
Then a simple power counting analysis gives,
\begin{equation}
{df^{\rm diff}_{g/A}(\beta x_\pom; x_\pom,t)\over dx_\pom\,dt}
\sim
{{\rm const.} \over (1-\beta) x_\pom^2}\
\int_0^\infty dq_T^2\
{\theta(q_T^2 > (1-\beta)\,x_\pom\,m^2) \over
 \{q_T^2/[(1-\beta)\,x_\pom]\}^2}
\end{equation}
for $x_\pom \ll 1$ and $(1-\beta) \ll 1$. Performing the integral
gives
\begin{equation}
\frac{df^{\rm diff}_{g/A}(\beta x_\pom; x_\pom,t)}{dx_\pom\,dt}
\sim (1-\beta)^0 x_{\pom}^{-1}.
\label{dfga}
\end{equation}
This agrees with our numerical findings.  We find it reassuring that
a fully consistent field theoretic calculation of the diffractive
gluon distribution gives results that agree with expectations similar
to our analytic arguments in Sec.~\ref{TauGlue}.


\end{document}